\documentclass[prd,aps,showpacs,epsf,floats,onecolumn]{revtex4}%
\usepackage{amssymb}
\usepackage{amsfonts}
\usepackage{amsmath}
\usepackage{graphicx}%
\providecommand{\U}[1]{\protect\rule{.1in}{.1in}}

\begin{document}
\title{\textbf{Theoretical analysis of a nearly optimal analog quantum search}}
\author{\textbf{Carlo Cafaro}$^{1}$ and \textbf{Paul M.\ Alsing}$^{2}$}
\affiliation{$^{1}$SUNY Polytechnic Institute, 12203 Albany, New York, USA}
\affiliation{$^{2}$Air Force Research Laboratory, Information Directorate, 13441 Rome, New
York, USA}

\begin{abstract}
We analyze the possibility of modifying the original Farhi-Gutmann Hamiltonian
algorithm in order to speed up the procedure for producing a suitably
distributed unknown normalized quantum mechanical state. Such a modification
is feasible provided only a nearly optimal fidelity is sought. We propose to
select the lower bounds of the nearly optimal fidelity values such that their
deviations from unit fidelity are less than the minimum error probability
characterizing the optimum ambiguous discrimination scheme between the two
nonorthogonal quantum states yielding the chosen nearly optimal fidelity
values. Departing from the working assumptions of perfect state overlap and
uniform distribution of the target state on the unit sphere in $N$-dimensional
complex Hilbert space, we determine that the modified algorithm can indeed
outperform the original analog counterpart of a quantum search algorithm. This
performance enhancement occurs in terms of speed for a convenient choice of
both the ratio $\gamma=E^{\prime}/E$ between the energy eigenvalues
$E^{\prime}$ and $E$ of the modified search Hamiltonian and the quantum
mechanical overlap $x$ between the source and the target states. Finally, we
briefly discuss possible analytical improvements of our investigation together
with its potential relevance in practical quantum engineering applications.

\end{abstract}

\maketitle

\bigskip\pagebreak

\section{Introduction}

In 1997, Grover presented a quantum algorithm for solving very large database
search problems \cite{grover97}, one of the outstanding problems in
computational science \cite{knuth75}. Grover's search algorithm enables
searching for an unknown marked item in an unstructured database of $N$ items
by accessing the database a minimum number of times. From a classical
standpoint, one needs to test $N/2$ items, on average, before identifying the
correct item. With Grover's algorithm however, the same task can be completed
successfully with a complexity of order $\sqrt{N}$, that is, with a quadratic
speed up. The original formulation of Grover's algorithm was presented in
terms of a discrete sequence of unitary logic gates (digital quantum
computation). Alternatively, Farhi and Gutmann proposed an analog version of
Grover's algorithm in Ref. \cite{farhi98} where the state of the quantum
register evolves continuously in time under the action of a suitably chosen
driving Hamiltonian (analog quantum computation). For recent discussions on
the transition from the digital to analog quantum computational setting for
Grover's algorithm, we refer to Ref. \cite{carlo1,carlo2, cafaro2017}. The
formulation of the analog algorithm proposed by Farhi and Gutmann can be
briefly described as follows. Consider $N$ mutually orthonormal states and
assume that only one of them has non-zero energy (energetically marked state).
Starting from an equal superposition of the $N$ states, the problem is to
determine how quickly the system can evolve under a given Hamiltonian so that
the energetically marked state is reached with certainty. Within the working
assumption of time-independent Hamiltonian evolutions, Farhi and Gutmann
showed that their algorithm required a minimum time of the order $\sqrt{N}$,
thus yielding the same complexity as Grover's original algorithm.
Interestingly, the evolution time for the analog algorithm scaled like
$\sqrt{N}/E$, with $E$ denoting the energy of the marked state, just as
demanded by the time-energy uncertainty principle \cite{mandelstam45}. In
Refs. \cite{mandelstam45,vaidman92,uffink93}, it was shown that the minimum
time interval required for an isolated quantum system to evolve to an
orthogonal (that is, perfectly distinguishable) state is inversely
proportional to its constant energy spread. Such minimum time is also known as
orthogonalization time in the literature. In Ref. \cite{margolus98}, the
maximum speed of evolution of an isolated quantum system was interpreted in
terms of the maximum number of distinct states that the system can explore in
a fixed time interval. This fixed time interval was shown to be inversely
proportional to the system's quantum mechanical average energy minus its
ground state energy. For a unified tight bound on the rate of quantum
dynamical evolution based on both the average energy and the energy spread of
the system, we refer to Ref. \cite{levitin09}. The presence of a finite
non-orthogonality in the context of quantum limits to dynamical evolutions was
considered, for instance, in Ref. \cite{trifonov99} and Ref.
\cite{giovannetti03}. In the former work, the expression of the states that
minimize the time necessary to evolve, under a time-independent Hamiltonian,
from an initial quantum state to a final state that is only approximately
orthogonal was found. It is worth noting that the final state is only
approximately orthogonal for a given upper bound on the mean energy of a
general closed system. In the latter work, on the contrary, an extension of
the concept of quantum speed limit to mixed states in the case where the
quantum system does not evolve from an initial state to an orthogonal state
(that is, the case of non-zero fidelity) was proposed.

In our paper, instead of allowing for a finite non-orthogonality between the
quantum states that leads to a non-zero fidelity, we consider a complementary
scenario. We allow for a finite non-parallelism that yields a non-unit
fidelity. The consideration of nearly optimal state searching is instructive
for various reasons. First, it is reasonable to take into account the fact
that actual measurements always involve a countable set of outcomes and thus,
any realistic measurement device has finite resolution \cite{partovi}%
.\ Second, finite resolution measurements can reveal more about the physical
properties of quantum systems than precise measurements \cite{furusawa}.
Third, it is known that a quantum measurement can enhance the transition
probability between two pure states \cite{fritz10} and, finally, it is
indisputable that quantum measurement cannot perfectly discriminate between
two nonorthogonal pure states \cite{chefles00}.

Motivated by these considerations, the questions we address in this paper can
be outlined as follows: First, what is the minimum time necessary to reach a
given nearly optimal fidelity value in a continuous time quantum search
algorithm? Second, is it possible to formulate a modified version of the
original Farhi-Gutmann search algorithm that leads to such a selected fidelity
value in a shorter time than the one required by the original Farhi-Gutmann
algorithm? Third, how can we reasonably justify the choice of a lower bound
for such a nearly optimal fidelity value? We demonstrate that it is possible
to extend the original work performed by Farhi and Gutmann in order to speed
up the procedure for producing a suitably distributed unknown normalized
quantum state with high non-unit fidelity. In particular, by modifying the
original working assumptions of perfect state overlap and uniform distribution
of the target state on the unit sphere in the $N$-dimensional complex Hilbert
space, the newly proposed algorithm can indeed outperform the original analog
counterpart of a quantum search algorithm. This is achieved for a convenient
choice of both\textbf{ }the ratio\textbf{ }$\gamma\overset{\text{def}}%
{=}E^{\prime}/E$ between the eigenvalues $E^{\prime}$ and $E$ of the modified
search Hamiltonian and the quantum mechanical overlap $x$ between the initial
and target states.

The layout of the remainder of this paper is as follows. In Section II, we
extend the Farhi-Gutmann optimality proof to the case of an imperfect quantum
state overlap. In Section III, we present the transition probability for the
modified version of the Farhi-Gutmann analog algorithm. In Section IV, we
explain how we choose the lower bounds of the nearly optimal fidelity values
by means of quantum discrimination techniques. Specifically, we impose that
deviations from unit fidelity are less than the minimum error probability
characterizing the optimum ambiguous discrimination scheme between the two
nonorthogonal quantum states yielding the selected nearly optimal fidelity
values. In Section V, we compare in terms of minimum run time the original and
the modified quantum search algorithms and identify a two-dimensional
parametric region where the modified algorithm outperforms the original one
under the working assumption of imperfect state transfer compatible with
finite precision quantum measurements. Our concluding remarks appear in
Section VI. Finally, technical details on the numerical values of the quantum
overlap $x$ can be found in Appendix A.

\section{The modified optimality proof}

In this section, we modify the original optimality proof by Farhi and Gutmann
in Ref. \cite{farhi98} by relaxing the requirement of finding the desired
target state with certainty. In particular, we wish to determine how the
search time shortens if we only want to achieve a nearly optimal fidelity value.

The optimality proof provided by Farhi and Gutmann in Ref. \cite{farhi98} for
producing the target state $\left\vert w\right\rangle $ relies on two key
working assumptions. First, the target state is produced with absolute
certainty. In other words, the modulus squared of the quantum state overlap,
that is, the transition probability or transition fidelity at the end of the
production procedure is equal to unity. Second, the target state $\left\vert
w\right\rangle $ is an unknown element of a given orthonormal basis $\left\{
\left\vert a\right\rangle \right\}  $ with $1\leq a\leq N$ of an
$N$-dimensional complex Hilbert space $\mathcal{H}_{2}^{n}$ with
$N\overset{\text{def}}{=}2^{n}$. Our modified optimality proof for an
imperfect quantum state overlap follows closely the lines of the original
optimality proof presented in Ref. \cite{farhi98}. Consider a quantum system,
assumed to be initially in a state $\left\vert i\right\rangle $ at time $t=0$
that does not depend on the target state $\left\vert w\right\rangle $ and
evolves by means of a Hamiltonian $\mathcal{H}$ given by,%
\begin{equation}
\mathcal{H}\overset{\text{def}}{=}\mathcal{H}_{w}+\mathcal{H}_{D}\text{,}
\label{hamo}%
\end{equation}
where $\mathcal{H}_{w}\overset{\text{def}}{=}E\left\vert w\right\rangle
\left\langle w\right\vert $. For the discussion that follows, it is
unnecessary to specify the explicit expression for the driving Hamiltonian
$\mathcal{H}_{D}$ in Eq. (\ref{hamo}). The problem being investigated is that
of finding a lower bound $\tilde{t}$ on the time $t$ for the $\left\vert
w\right\rangle $-independent state $\left\vert i\right\rangle $ to evolve into
the state $\left\vert w\right\rangle $ given that its evolution is governed by
the time-independent Hamiltonian $\mathcal{H}$ in Eq. (\ref{hamo}). Let us
consider the following two initial value problems,%
\begin{equation}
\left\{
\begin{array}
[c]{c}%
i\hslash\frac{d}{dt}\left\vert \psi_{w}\left(  t\right)  \right\rangle
=\mathcal{H}\left\vert \psi_{w}\left(  t\right)  \right\rangle \\
\left\vert \psi_{w}\left(  0\right)  \right\rangle =\left\vert i\right\rangle
\end{array}
\right.  \text{, and }\left\{
\begin{array}
[c]{c}%
i\hslash\frac{d}{dt}\left\vert \psi\left(  t\right)  \right\rangle
=\mathcal{H}_{D}\left\vert \psi\left(  t\right)  \right\rangle \\
\left\vert \psi\left(  0\right)  \right\rangle =\left\vert i\right\rangle
\end{array}
\right.  \text{,}%
\end{equation}
where $\hslash$ denotes the reduced Planck constant. In what follows, we
depart from Farhi and Gutmann original optimality proof. We do not require
achieving the absolute certainty of producing the target state at the end of
the evolution and assume that at time\textbf{ }$\tilde{t}$\textbf{ }we have,%
\begin{equation}
\left\vert \psi_{w}\left(  \tilde{t}\right)  \right\rangle \overset
{\text{def}}{=}\left\langle w|\psi_{w}\left(  \tilde{t}\right)  \right\rangle
\left\vert w\right\rangle +\left\langle w_{\bot}|\psi_{w}\left(  \tilde
{t}\right)  \right\rangle \left\vert w_{\bot}\right\rangle \text{.}
\label{imperfect}%
\end{equation}
The normalization condition for\textbf{ }$\left\vert \psi_{w}\left(  \tilde
{t}\right)  \right\rangle $ requires that the probability amplitudes satisfy
the following constraint equation,%
\begin{equation}
\left\vert \left\langle w|\psi_{w}\left(  \tilde{t}\right)  \right\rangle
\right\vert ^{2}+\left\vert \left\langle w_{\bot}|\psi_{w}\left(  \tilde
{t}\right)  \right\rangle \right\vert ^{2}=1\text{.}%
\end{equation}
The quantity $\left\vert w_{\bot}\right\rangle $\textbf{ }denotes a
(normalized) state orthogonal to the target state\textbf{ }$\left\vert
w\right\rangle $\textbf{ }contained in the two-dimensional subspace of the
full Hilbert space spanned by $\left\vert w\right\rangle $ and\textbf{
}$\left\vert w_{\bot}\right\rangle $\textbf{. }A suitable choice for the
parametrization of the state\textbf{ }$\left\vert \psi_{w}\left(  \tilde
{t}\right)  \right\rangle $\textbf{ }in Eq. (\ref{imperfect}) is given by,%
\begin{equation}
\left\vert \psi_{w}\left(  \tilde{t}\right)  \right\rangle =\cos\left(
\delta\right)  \left\vert w\right\rangle +e^{-i\left(  \theta_{w}%
-\theta_{w_{\bot}}\right)  }\sin\left(  \delta\right)  \left\vert w_{\bot
}\right\rangle \text{,}%
\end{equation}
where\textbf{ }$\theta_{w}$\textbf{ }and\textbf{ }$\theta_{w_{\bot}}$\textbf{
}are arbitrary \emph{real} phases\textbf{ }while\textbf{ }$\delta$\textbf{ }is
assumed to be a small and positive \emph{real} parameter. Finally, for
convenience and without loss of generality, we set $\theta_{w}-\theta
_{w_{\bot}}=\pi/2$ so that $\left\vert \psi_{w}\left(  \tilde{t}\right)
\right\rangle $ becomes%
\begin{equation}
\left\vert \psi_{w}\left(  \tilde{t}\right)  \right\rangle =\cos\left(
\delta\right)  \left\vert w\right\rangle -i\sin\left(  \delta\right)
\left\vert w_{\bot}\right\rangle \text{,} \label{imperfect2}%
\end{equation}
where $i$ denotes the \emph{complex} imaginary unit. In what follows, we
assume\textbf{ }$0\leq\delta\ll1$\textbf{ }so that\textbf{ }$\cos^{2}\left(
\delta\right)  =1-\delta^{2}+\mathcal{O}\left(  \delta^{4}\right)  $\textbf{
}and\textbf{ }$\sin^{2}\left(  \delta\right)  =\delta^{2}+\mathcal{O}\left(
\delta^{4}\right)  $\textbf{. }For the sake of clarity, we point out that the
validity of this latter working assumption will be explicitly clarified by
means of Eq. (\ref{deltaequation}) and Fig. $2$ in the next section\textbf{.
}Furthermore, we have%
\begin{equation}
\cos\left(  \delta\right)  =1-(1/2)\delta^{2}+\mathcal{O}\left(  \delta
^{4}\right)  \text{, and }\sin\left(  \delta\right)  =\delta+\mathcal{O}%
\left(  \delta^{3}\right)  \text{.} \label{imperfect3}%
\end{equation}
To find a lower bound $\tilde{t}$ on the time $t$, we proceed as follows.
First, consider the following quantity%
\begin{equation}
\sum_{w=1}^{N}\left\Vert \left\vert \psi_{w}\left(  \tilde{t}\right)
\right\rangle -\left\vert \psi\left(  \tilde{t}\right)  \right\rangle
\right\Vert ^{2}\text{.} \label{sum}%
\end{equation}
By using Eqs. (\ref{imperfect2}) and (\ref{imperfect3}) in the expression for
$\left\Vert \left\vert \psi_{w}\left(  \tilde{t}\right)  \right\rangle
-\left\vert \psi\left(  \tilde{t}\right)  \right\rangle \right\Vert ^{2}$
appearing in Eq. (\ref{sum}), we obtain after some algebra,
\begin{equation}
\left\Vert \left\vert \psi_{w}\left(  \tilde{t}\right)  \right\rangle
-\left\vert \psi\left(  \tilde{t}\right)  \right\rangle \right\Vert
^{2}=2-\cos\left(  \delta\right)  \left[  \left\langle \psi\left(  \tilde
{t}\right)  |w\right\rangle +\left\langle w|\psi\left(  \tilde{t}\right)
\right\rangle \right]  +i\sin\left(  \delta\right)  \left[  \left\langle
\psi\left(  \tilde{t}\right)  |w_{\perp}\right\rangle -\left\langle w_{\perp
}|\psi\left(  \tilde{t}\right)  \right\rangle \right]  \text{.} \label{just}%
\end{equation}
Substituting Eq. (\ref{just}) into Eq. (\ref{sum}), we obtain%
\begin{equation}
\sum_{w=1}^{N}\left\Vert \left\vert \psi_{w}\left(  \tilde{t}\right)
\right\rangle -\left\vert \psi\left(  \tilde{t}\right)  \right\rangle
\right\Vert ^{2}=2N-\cos\left(  \delta\right)  \sum_{w=1}^{N}\left[
\left\langle \psi\left(  \tilde{t}\right)  |w\right\rangle +\left\langle
w|\psi\left(  \tilde{t}\right)  \right\rangle \right]  +i\sin\left(
\delta\right)  \sum_{w=1}^{N}\left[  \left\langle \psi\left(  \tilde
{t}\right)  |w_{\perp}\right\rangle -\left\langle w_{\perp}|\psi\left(
\tilde{t}\right)  \right\rangle \right]  \text{.} \label{omega1}%
\end{equation}
Observe that for a set of \emph{complex }amplitude coefficients $\left\{
c_{j}\right\}  $ with $1\leq j\leq N$, we find%
\begin{equation}
\sum_{j=1}^{N}\left\vert c_{j}\right\vert ^{2}=1\Rightarrow\sum_{j=1}%
^{N}\left\vert c_{j}\right\vert \leq N^{\frac{1}{2}}\text{.} \label{omega2}%
\end{equation}
Therefore, by exploiting Eq. (\ref{omega2}), we\textbf{ }observe that%
\begin{equation}
\sum_{w=1}^{N}\left\Vert \left\vert \psi_{w}\left(  \tilde{t}\right)
\right\rangle -\left\vert \psi\left(  \tilde{t}\right)  \right\rangle
\right\Vert ^{2}\geq2N-2\sqrt{N}\cos\left(  \delta\right)  -2\sqrt{N}%
\sin\left(  \delta\right)  \text{.} \label{vafa}%
\end{equation}
By using Eq. (\ref{imperfect3}) in Eq. (\ref{vafa}) and neglecting second
order terms in $\delta$ we arrive at,%
\begin{equation}
\sum_{w=1}^{N}\left\Vert \left\vert \psi_{w}\left(  \tilde{t}\right)
\right\rangle -\left\vert \psi\left(  \tilde{t}\right)  \right\rangle
\right\Vert ^{2}\geq\left(  2N-2\sqrt{N}\right)  \left(  1-\delta\right)
+2\left(  N-2\sqrt{N}\right)  \delta\text{.} \label{vafa3}%
\end{equation}
Observe that for $N\geq4$, we have both $2N-2\sqrt{N}\geq N$ and $N-2\sqrt
{N}\geq0$. Finally, Eq. (\ref{vafa3}) reduces to,%
\begin{equation}
\sum_{w=1}^{N}\left\Vert \left\vert \psi_{w}\left(  \tilde{t}\right)
\right\rangle -\left\vert \psi\left(  \tilde{t}\right)  \right\rangle
\right\Vert ^{2}\geq N\left(  1-\delta\right)  \text{.} \label{prima}%
\end{equation}
Now, let us consider the following expression,%
\begin{equation}
\frac{d}{dt}\left\Vert \left\vert \psi_{w}\left(  t\right)  \right\rangle
-\left\vert \psi\left(  t\right)  \right\rangle \right\Vert ^{2}\text{.}%
\end{equation}
After some algebra and exploiting the normalization conditions $\left\langle
\psi_{w}\left(  t\right)  |\psi_{w}\left(  t\right)  \right\rangle =1$ and
$\left\langle \psi\left(  t\right)  |\psi\left(  t\right)  \right\rangle =1$,
we obtain%
\begin{equation}
\left\Vert \left\vert \psi_{w}\left(  t\right)  \right\rangle -\left\vert
\psi\left(  t\right)  \right\rangle \right\Vert ^{2}=2-\left\langle \psi
_{w}\left(  t\right)  |\psi\left(  t\right)  \right\rangle -\left\langle
\psi\left(  t\right)  |\psi_{w}\left(  t\right)  \right\rangle \text{.}
\label{norma2}%
\end{equation}
Differentiating with respect to time the LHS (left-hand-side) and the RHS
(right-hand-side) of Eq. (\ref{norma2}) and recalling that $z+z^{\ast
}=2\operatorname{Re}\left(  z\right)  $ for any $z\in%
\mathbb{C}
$ where $z^{\ast}$ denotes the complex conjugate of $z$, we find%
\begin{equation}
\frac{d}{dt}\left\Vert \left\vert \psi_{w}\left(  t\right)  \right\rangle
-\left\vert \psi\left(  t\right)  \right\rangle \right\Vert ^{2}%
=-2\operatorname{Re}\left\{  \frac{d}{dt}\left[  \left\langle \psi_{w}\left(
t\right)  |\psi\left(  t\right)  \right\rangle \right]  \right\}  \text{,}%
\end{equation}
that is,%
\begin{equation}
\frac{d}{dt}\left\Vert \left\vert \psi_{w}\left(  t\right)  \right\rangle
-\left\vert \psi\left(  t\right)  \right\rangle \right\Vert ^{2}%
=-2\operatorname{Re}\left[  \left\langle \frac{d\left\vert \psi_{w}\left(
t\right)  \right\rangle }{dt}|\psi\left(  t\right)  \right\rangle
+\left\langle \psi_{w}\left(  t\right)  |\frac{d\left\vert \psi\left(
t\right)  \right\rangle }{dt}\right\rangle \right]  \text{.} \label{omega5}%
\end{equation}
Recalling that $i\hslash\frac{d}{dt}\left\vert \psi_{w}\left(  t\right)
\right\rangle =\mathcal{H}\left\vert \psi_{w}\left(  t\right)  \right\rangle $
and $i\hslash\frac{d}{dt}\left\vert \psi\left(  t\right)  \right\rangle
=\mathcal{H}_{D}\left\vert \psi\left(  t\right)  \right\rangle $ and observing
that $\operatorname{Re}\left(  iz\right)  =-\operatorname{Im}\left(  z\right)
$ for any $z\in%
\mathbb{C}
$, Eq. (\ref{omega5}) becomes%
\begin{equation}
\frac{d}{dt}\left\Vert \left\vert \psi_{w}\left(  t\right)  \right\rangle
-\left\vert \psi\left(  t\right)  \right\rangle \right\Vert ^{2}=\frac
{2}{\hslash}\operatorname{Im}\left[  \left\langle \psi_{w}\left(  t\right)
|\mathcal{H}_{w}|\psi\left(  t\right)  \right\rangle \right]  \text{.}%
\end{equation}
Employing the Cauchy-Schwarz inequality and recalling that $\left\langle
\psi_{w}\left(  t\right)  |\psi_{w}\left(  t\right)  \right\rangle =1$, we
obtain%
\begin{align}
\frac{d}{dt}\left\Vert \left\vert \psi_{w}\left(  t\right)  \right\rangle
-\left\vert \psi\left(  t\right)  \right\rangle \right\Vert ^{2}  &  =\frac
{2}{\hslash}\operatorname{Im}\left[  \left\langle \psi_{w}\left(  t\right)
|\mathcal{H}_{w}|\psi\left(  t\right)  \right\rangle \right] \nonumber\\
& \nonumber\\
&  \leq\frac{2}{\hslash}\left\vert \left\langle \psi_{w}\left(  t\right)
|\mathcal{H}_{w}|\psi\left(  t\right)  \right\rangle \right\vert \nonumber\\
& \nonumber\\
&  \leq\frac{2}{\hslash}\left\Vert \left\vert \psi_{w}\left(  t\right)
\right\rangle \right\Vert \cdot\left\Vert \mathcal{H}_{w}\left\vert
\psi\left(  t\right)  \right\rangle \right\Vert =\frac{2}{\hslash}\left\Vert
\mathcal{H}_{w}\left\vert \psi\left(  t\right)  \right\rangle \right\Vert
\text{.} \label{inequal}%
\end{align}
Summing over the index $w$ in Eq. (\ref{inequal}), we have%
\begin{equation}
\frac{d}{dt}\left(  \sum_{w=1}^{N}\left\Vert \left\vert \psi_{w}\left(
t\right)  \right\rangle -\left\vert \psi\left(  t\right)  \right\rangle
\right\Vert ^{2}\right)  \leq\frac{2}{\hslash}\sum_{w=1}^{N}\left\Vert
\mathcal{H}_{w}\left\vert \psi\left(  t\right)  \right\rangle \right\Vert
\text{.} \label{aa22}%
\end{equation}
Since $\mathcal{H}_{w}\overset{\text{def}}{=}E\left\vert w\right\rangle
\left\langle w\right\vert $ and $\left\langle w|w\right\rangle =1$, and using
Eq. (\ref{omega2}), the RHS in the inequality written in Eq. (\ref{aa22})
becomes%
\begin{equation}
\frac{2}{\hslash}\sum_{w=1}^{N}\left\Vert \mathcal{H}_{w}\left\vert
\psi\left(  t\right)  \right\rangle \right\Vert =\frac{2E}{\hslash}\sum
_{w=1}^{N}\left\vert \left\langle w|\psi\left(  t\right)  \right\rangle
\right\vert \leq\frac{2E}{\hslash}\sqrt{N}\text{.} \label{aa23}%
\end{equation}
Combining Eqs. (\ref{aa22}) and (\ref{aa23}), we obtain%
\begin{equation}
\frac{d}{dt}\left(  \sum_{w=1}^{N}\left\Vert \left\vert \psi_{w}\left(
t\right)  \right\rangle -\left\vert \psi\left(  t\right)  \right\rangle
\right\Vert ^{2}\right)  \leq\frac{2E}{\hslash}\sqrt{N}\text{.} \label{tobe}%
\end{equation}
Integrating Eq. (\ref{tobe}) with respect to time and recalling the initial
conditions $\left\vert \psi_{w}\left(  0\right)  \right\rangle =\left\vert
\psi\left(  0\right)  \right\rangle =\left\vert i\right\rangle $, we find%
\begin{equation}
\sum_{w=1}^{N}\left\Vert \left\vert \psi_{w}\left(  t\right)  \right\rangle
-\left\vert \psi\left(  t\right)  \right\rangle \right\Vert ^{2}\leq\frac
{2E}{\hslash}\sqrt{N}t\text{.} \label{seconda}%
\end{equation}
Finally, by combining Eq. (\ref{prima}) and Eq. (\ref{seconda}) evaluated at
$t=\tilde{t}$, we have%
\begin{equation}
N\left(  1-\delta\right)  \leq\sum_{w=1}^{N}\left\Vert \left\vert \psi
_{w}\left(  \tilde{t}\right)  \right\rangle -\left\vert \psi\left(  \tilde
{t}\right)  \right\rangle \right\Vert ^{2}\leq\frac{2E}{\hslash}\sqrt{N}%
\tilde{t}\text{,}%
\end{equation}
that is,%
\begin{equation}
\frac{2E}{\hslash}\sqrt{N}\tilde{t}\geq N\left(  1-\delta\right)  \text{,}%
\end{equation}
or, equivalently,%
\begin{equation}
\tilde{t}\geq\mathbf{\ }\frac{\hslash}{2E}\left(  1-\delta\right)  \sqrt
{N}\text{.} \label{findingone}%
\end{equation}
Equation (\ref{findingone}) is our first finding reported in this paper and
can be regarded as the analog of Eq. (23) in Ref. \cite{trifonov99} obtained
in the context of finding states that minimize the evolution time to become an
\emph{almost} orthogonal state. We point out that although it is expected
that\textbf{ }$\tilde{t}_{\min}^{\left(  \text{approximate}\right)  }%
\leq\tilde{t}_{\min}^{\left(  \text{exact}\right)  }$\textbf{,} the scaling
behavior of\textbf{ }$\tilde{t}_{\min}^{\left(  \text{exact}\right)  }%
$\textbf{ }as a function of the angle\textbf{ }$\delta$\textbf{ }with\textbf{
}$0\leq\delta\ll1$\textbf{ }is not at all obvious. We emphasize that the
validity of Eq. (\ref{findingone}) is limited to the case in which the target
state $\left\vert w\right\rangle $ is an unknown element of a given
orthonormal basis $\left\{  \left\vert a\right\rangle \right\}  $ with $1\leq
a\leq N$ of an $N$-dimensional complex Hilbert space. We also recall that
$\delta$ is defined by the relation $\delta\overset{\text{def}}{=}\cos
^{-1}\left[  \left\langle w|\psi_{w}\left(  \tilde{t}\right)  \right\rangle
\right]  $ with\textbf{ }$\left\vert \psi_{w}\left(  t\right)  \right\rangle
\overset{\text{def}}{=}\hat{U}\left(  t\text{, }t_{0}\right)  \left\vert
\psi_{w}\left(  t_{0}\right)  \right\rangle $. The unitary operator $\hat
{U}\left(  t\text{, }t_{0}\right)  \overset{\text{def}}{=}e^{-\frac{i}%
{\hslash}\mathcal{H}\left(  t-t_{0}\right)  }$ is the usual quantum mechanical
evolution operator. Therefore in general, the quantity $\delta$ can depend on
certain characteristics of the chosen time-independent search Hamiltonian
$\mathcal{H}$.

For the sake of clarity, we point out that our proposed modified optimality
proof is not meant to be used to determine the optimality of a modified
continuous time quantum search algorithm. The lower bound in Eq.
(\ref{findingone}) is only\textbf{ }useful for providing the optimality of the
scaling of the complexity of any algorithm specified by the Hamiltonian in Eq.
(\ref{hamo}). In particular, its purpose is to find out the behavior with
respect to the parameter $\delta$ of the minimum time necessary to achieve a
desired nearly optimal transition probability value (that is, $\cos^{2}\left(
\delta\right)  $ with $0\leq\delta\ll1$) when evolving from a source state
$\left\vert i\right\rangle $ to an approximate target state $\left\vert
\psi_{w}\left(  \tilde{t}\right)  \right\rangle $ under the Hamiltonian in Eq.
(\ref{hamo}). In other words, we have not compared the optimality of two
distinct algorithms that aim to solve the same identical task. Instead, we are
interested in finding out how the performance of a single algorithm,
quantified in terms of the minimum time necessary to accomplish the task, is
affected when two distinct tasks have to be completed. In our case, the first
task (that is, the original task) is to find the target state with certainty
(that is, success probability equal to one). The second task (that is, the
modified task), instead, is to search for an approximate target state that
corresponds to a success probability that is only nearly equal to one. In
summary, when considering the original task, we go from $\left\vert
i\right\rangle $ to $\left\vert w\right\rangle $ with success probability one
and $\tilde{t}_{\min}^{\left(  \text{exact}\right)  }\overset{\text{def}}%
{=}(\hslash/2E)\sqrt{N}$. In the modified task, we go from $\left\vert
i\right\rangle $ to $\left\vert \psi_{w}\left(  \tilde{t}\right)
\right\rangle $ with success probability $\cos^{2}\left(  \delta\right)  $
where $0\leq\delta\ll1$ and $\tilde{t}_{\min}^{\left(  \text{approximate}%
\right)  }\overset{\text{def}}{=}(\hslash/2E)\left(  1-\delta\right)  \sqrt
{N}$. In both cases, the same algorithm described by the Hamiltonian in Eq.
(\ref{hamo}) is employed.

Our proposed modified optimality proof allowed us to formally introduce the
parameter $\delta$ in a general physical setting where it is not necessary to
specify the explicit expression of the driving Hamiltonian $\mathcal{H}_{D}$
in Eq. (\ref{hamo}). The following section, instead, includes a more specific
discussion on the functional form of the parameter $\delta$ for a specific
choice of the driving Hamiltonian.

\section{The modified Farhi-Gutmann search Hamiltonian}

In this section, we present the transition probability for the modified
version of the Farhi-Gutmann analog search algorithm. Assume that the
time-independent Hamiltonian $\mathcal{H}$ in Eq. (\ref{hamo}) is given by,%
\begin{equation}
\mathcal{H}\overset{\text{def}}{=}\mathcal{H}_{w}+\mathcal{H}_{D}=E\left\vert
w\right\rangle \left\langle w\right\vert +E^{\prime}\left\vert s\right\rangle
\left\langle s\right\vert \text{,} \label{hamilton2}%
\end{equation}
where,%
\begin{equation}
\mathcal{H}_{w}\overset{\text{def}}{=}E\left\vert w\right\rangle \left\langle
w\right\vert \text{ and, }\mathcal{H}_{D}\overset{\text{def}}{=}E^{\prime
}\left\vert s\right\rangle \left\langle s\right\vert \text{.}%
\end{equation}
Observe that when $E^{\prime}=E$, the Hamiltonian $\mathcal{H}$ reduces to the
Hamiltonian as originally proposed by Farhi and Gutmann in Ref. \cite{farhi98}%
. Furthermore, assume that $\left\vert w\right\rangle $ is the normalized
target state while $\left\vert s\right\rangle $ is the normalized initial
state with $\left\langle s|w\right\rangle =x$ that evolves in an unitary
fashion according to Schr\"{o}dinger's quantum mechanical evolution law,%
\begin{equation}
\left\vert s\right\rangle \mapsto e^{-\frac{i}{\hslash}\mathcal{H}t}\left\vert
s\right\rangle \text{,}%
\end{equation}
where the time-independent Hamiltonian $\mathcal{H}$ is given in Eq.
(\ref{hamilton2}). We wish to compute the time $t^{\ast}$ such that the
transition probability $\mathcal{P}\left(  t\right)  $ defined as,%
\begin{equation}
\mathcal{P}\left(  t\right)  \overset{\text{def}}{=}\left\vert \left\langle
w|e^{-\frac{i}{\hslash}\mathcal{H}t}|s\right\rangle \right\vert ^{2}\text{,}
\label{fidelity}%
\end{equation}
\textbf{ }assumes its maximum value\textbf{ }$\mathcal{P}\left(  t^{\ast
}\right)  =\mathcal{P}_{\max}$. Introducing the orthonormal basis\textbf{
}$\left\{  \left\vert w\right\rangle \text{, }\left\vert r\right\rangle
\right\}  $ with $\left\vert r\right\rangle \overset{\text{def}}{=}\left(
1-x^{2}\right)  ^{-\frac{1}{2}}\left(  \left\vert s\right\rangle -x\left\vert
w\right\rangle \right)  $ and $\left\vert s\right\rangle \overset{\text{def}%
}{=}x\left\vert w\right\rangle +\sqrt{1-x^{2}}\left\vert r\right\rangle $,
after some straightforward but tedious matrix algebra computations, the
explicit expression\textbf{ }$\mathcal{P}\left(  t\text{; }x\text{, }E\text{,
}E^{\prime}\right)  $\textbf{ }of\textbf{ }$\mathcal{P}\left(  t\right)
$\textbf{ }in\ Eq. (\ref{fidelity}) becomes%
\begin{equation}
\mathcal{P}\left(  t\text{; }x\text{, }E\text{, }E^{\prime}\right)
=\frac{x^{2}\left(  1+\gamma\right)  ^{2}}{4x^{2}\gamma+\left(  1-\gamma
\right)  ^{2}}\sin^{2}\left(  \frac{1}{2\hslash}\sqrt{4x^{2}EE^{\prime
}+\left(  E^{\prime}-E\right)  ^{2}}t\right)  +x^{2}\cos^{2}\left(  \frac
{1}{2\hslash}\sqrt{4x^{2}EE^{\prime}+\left(  E^{\prime}-E\right)  ^{2}%
}t\right)  \text{,} \label{EF}%
\end{equation}
where $\gamma\overset{\text{def}}{=}E^{\prime}/E$ in Eq. (\ref{EF}). Observe
that for $E=E^{\prime}$, we recover from the expression of $\mathcal{P}\left(
t\text{; }x\text{, }E\text{, }E^{\prime}\right)  $ in Eq. (\ref{EF}) the
original fidelity expression as obtained by Farhi and Gutmann in Ref.
\cite{farhi98},%
\begin{equation}
\mathcal{P}_{\text{Farhi-Gutmann}}\left(  t\text{; }x\text{, }E\right)
=\sin^{2}\left(  \frac{Ex}{\hslash}t\right)  +x^{2}\cos^{2}\left(  \frac
{Ex}{\hslash}t\right)  \text{.} \label{34}%
\end{equation}
\begin{figure}[ptb]
\centering
\includegraphics[width=0.35\textwidth] {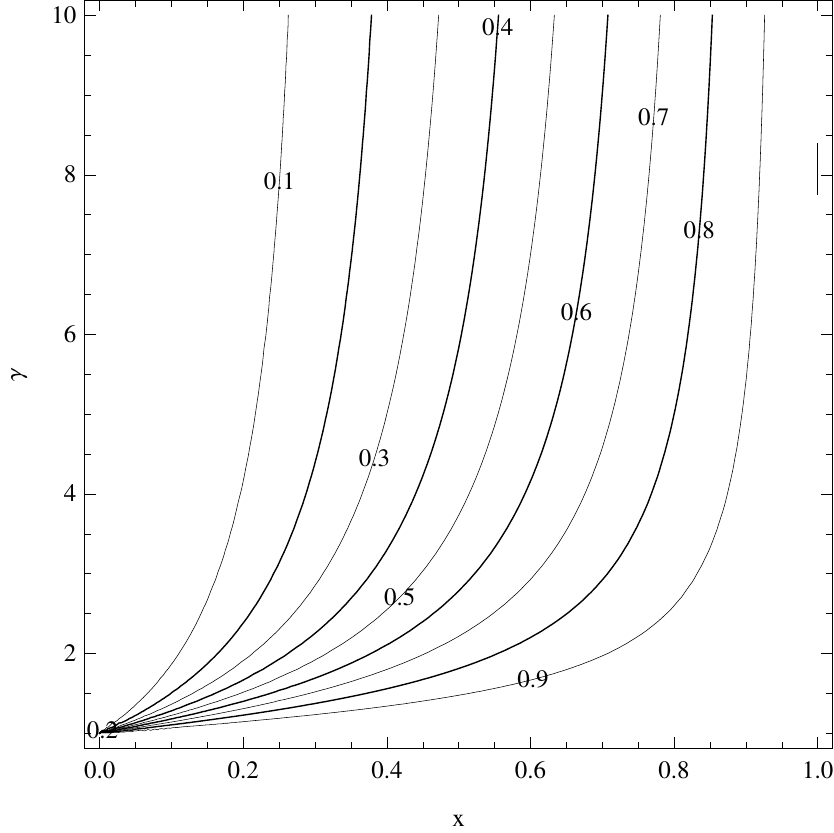}\caption{Contour plot exhibiting
$\gamma$ v.s. $x$ with $1\leq\gamma\leq10$ and $0\leq x\leq1$. The level
curves are given by $P_{g}^{\max}\left(  x\text{, }\gamma\right)  =c$ where
$c$ is a constant such that $0.1\leq c<1$.}%
\label{fig1}%
\end{figure}To find the instant $t^{\ast}$ for which $\mathcal{P}\left(
t^{\ast}\text{; }x\text{, }E\text{, }E^{\prime}\right)  =\mathcal{P}_{\max}$,
we proceed as follows. First, from Eq. (\ref{EF}), we observe that
$\mathcal{P}_{\max}$ equals
\begin{equation}
\mathcal{P}_{\max}\left(  x\text{, }\gamma\right)  =\frac{x^{2}\left(
1+\gamma\right)  ^{2}}{4x^{2}\gamma+\left(  1-\gamma\right)  ^{2}}\text{.}
\label{DDD}%
\end{equation}
In Fig. $1$, we present the contour plot that exhibits $\gamma$ v.s. $x$ with
$1\leq\gamma\leq10$ and $0\leq x<1$. In particular, the level curves are
defined by the relation $\mathcal{P}_{\max}\left(  x\text{, }\gamma\right)
=c$ (with $\mathcal{P}_{\max}\left(  x\text{, }\gamma\right)  =\mathcal{P}%
_{\text{g}}^{\max}$ in Eq.\ (\ref{gq})) where $c$ is a constant with numerical
values such that $0.1\leq c<1$. As a side remark, notice that when
$E=E^{\prime}$, that is $\gamma=1$, $\mathcal{P}_{\max}=1$ for any value of
$x$ in agreement with what was\textbf{ }reported in Ref. \cite{farhi98}.
Second, the instant $t^{\ast}$ for which $\mathcal{P}\left(  t^{\ast}\text{;
}x\text{, }E\text{, }E^{\prime}\right)  =\mathcal{P}_{\max}$ with
$\mathcal{P}\left(  t\text{; }x\text{, }E\text{, }E^{\prime}\right)  $ and
$\mathcal{P}_{\max}$ in Eqs. (\ref{EF}) and (\ref{DDD}), respectively,
satisfies the relation%
\begin{equation}
\frac{1}{2\hslash}\sqrt{4x^{2}EE^{\prime}+\left(  E^{\prime}-E\right)  ^{2}%
}t^{\ast}=\frac{\pi}{2}\text{,}%
\end{equation}
that is, since $E^{\prime}=\gamma E$,%
\begin{equation}
t^{\ast}\left(  x\text{, }\gamma\right)  \overset{\text{def}}{=}\frac
{\pi\hslash}{2E}\frac{2}{\sqrt{4x^{2}\gamma+\left(  1-\gamma\right)  ^{2}}%
}\text{.} \label{start}%
\end{equation}
Once again, observe that in the limiting case of $\gamma=1$, we recover the
original relation found in Ref. \cite{farhi98},%
\begin{equation}
t_{\text{Farhi-Gutmann}}^{\ast}\left(  x\right)  \overset{\text{def}}{=}%
\frac{\pi\hslash}{2E}\frac{1}{x}\text{.}%
\end{equation}
We recall that at the end of Section II, we stated that the quantity $\delta$
can depend on specific properties of the Hamiltonian $\mathcal{H}$\textbf{ }in
Eq. (\ref{hamilton2}). By using Eqs. (\ref{imperfect}) and (\ref{DDD}), we can
finally obtain the explicit functional form of $\delta$ in terms of the two
parameters $x$ and $\gamma$,%
\begin{equation}
\delta=\delta\left(  x\text{, }\gamma\right)  \overset{\text{def}}{=}\cos
^{-1}\left[  \frac{\left(  1+\gamma\right)  x}{\sqrt{\left(  1-\gamma\right)
^{2}+4\gamma x^{2}}}\right]  \text{.} \label{deltaequation}%
\end{equation}
Using Eq. (\ref{deltaequation}), we present in Fig. 2 a plot of $\delta\left(
x\right)  $ versus $x$ for a number of fixed values of $\gamma$ greater than
one. Finally, observe that for $\gamma=1$, $\delta\left(  x\text{, }%
\gamma\right)  =0$ for any choice of $x$ and we recover the limiting case
discussed in Ref. \cite{farhi98}.

\begin{figure}[ptb]
\centering
\includegraphics[width=0.4\textwidth] {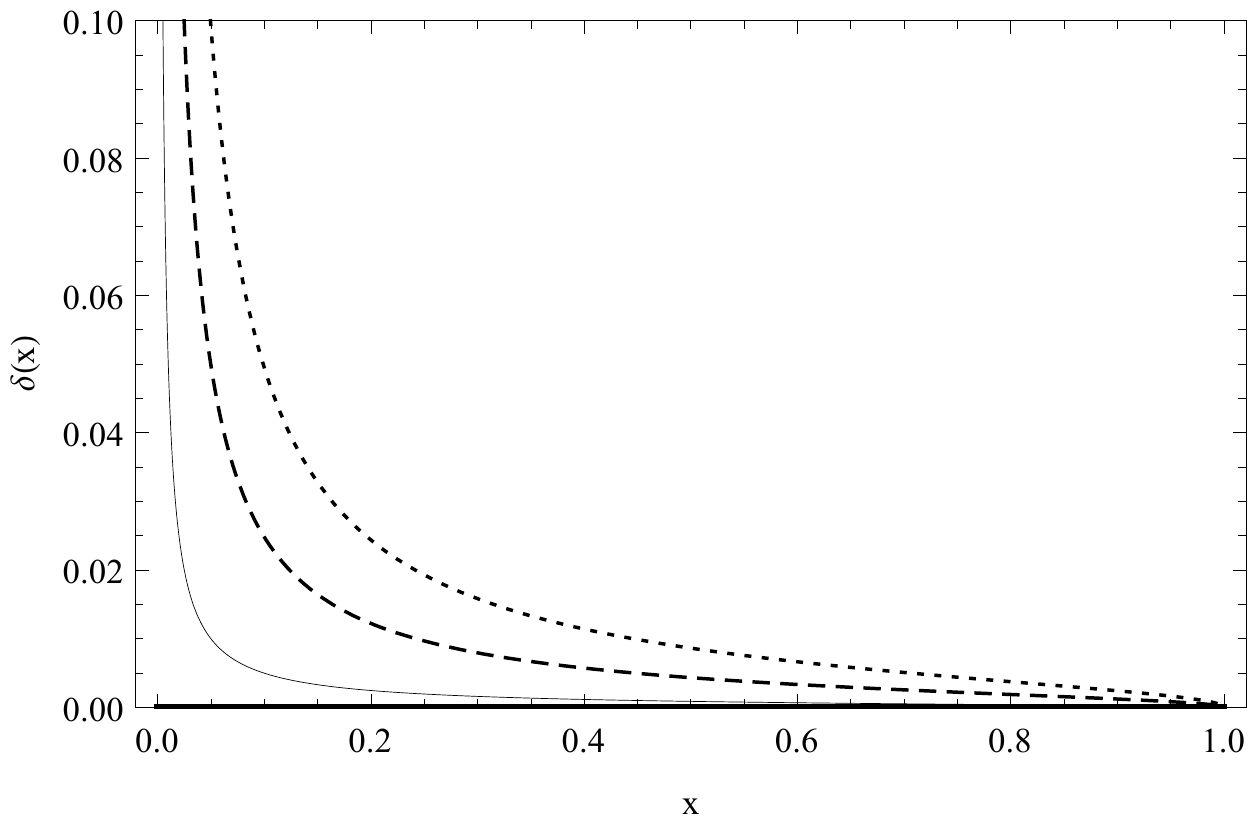}\caption{Plot of $\delta\left(
x\right)  $ v.s. $x$ assuming that $\gamma=1$ (thick line), $\gamma=1.001$
(thin line), $\gamma=1.010$ (dashed line), and $\gamma=1.1$ (dotted line).}%
\label{fig2}%
\end{figure}

\section{Lower bounds on the fidelity values}

Having introduced the parameter $\delta$ in a general setting in Section II
(see Eq. (\ref{findingone})) and in a more specific framework in Section III
(see Eq. (\ref{deltaequation})), we would like to address in\ Section IV the
following question:\ How big can $\delta$ be chosen? In principle, we could
arbitrarily choose a nearly optimal threshold value of the error probability
for quantum search. However, in what follows, we propose a way to justify to a
certain extent this arbitrariness with some plausible line of reasoning.
First, at the end of the search, we expect that the approximate and the exact
target states will be very close. Therefore, the error probability for quantum
search (that is, its deviation from the unit fidelity) will be very
small.\ Second, since these target states have a large overlap, it will be
difficult to distinguish them. Therefore, we also expect that the error
probability for quantum state discrimination will be large. Given these two
considerations, it is intuitive that in our working conditions the error
probability for quantum search will be smaller than the error probability for
quantum state discrimination. However, even in the presence of large overlaps
between quantum states, it is possible to device highly successful quantum
state discrimination protocols where the minimum error probability can be made
arbitrarily small. We propose to select the minimum success probability values
for quantum search by imposing that the deviation from the unit fidelity is
smaller than the smallest (compatible with realistic finite-precision quantum
measurements) minimum error probability achievable in highly successful
quantum state discrimination protocols for quantum states with very large overlaps.

If we desire to consider a very small departure\textbf{ }$\Delta
\mathcal{F}\left(  \delta\right)  $\textbf{ }from the unit fidelity,%
\begin{equation}
\Delta\mathcal{F}\left(  \delta\right)  \overset{\text{def}}{=}\mathcal{F}%
_{\text{optimal}}-\mathcal{F}_{\text{nearly-optimal}}\left(  \delta\right)
=1-\cos^{2}\left(  \delta\right)  \text{,} \label{40}%
\end{equation}
we expect\textbf{ }$0\leq1-\cos^{2}\left(  \delta\right)  \leq\varepsilon$
with $0\leq\varepsilon\ll1$\textbf{, }that is\textbf{ }%
\begin{equation}
0\leq\delta\leq\delta_{\max}\left(  \varepsilon\right)  \overset{\text{def}%
}{=}\cos^{-1}\left(  \sqrt{1-\varepsilon}\right)  \text{.} \label{epsilon}%
\end{equation}
Therefore, the question becomes: How small should the\textbf{ }\emph{real}%
\textbf{ }parameter\textbf{ }$\varepsilon$\textbf{ }in Eq. (\ref{epsilon}) be
chosen? How do we select this value? Can we physically motivate such a choice
for $\varepsilon$? In what follows, motivated by theoretical quantum state
discrimination principles \cite{chefles00,croke09} and acknowledging the
presence of imperfections in actual quantum measurement settings that limit
the precision of the measurement of angles \cite{gisin12,mario18}, we propose
a way to select the numerical upper (lower) bounds for $\delta$ \textbf{(}%
$\mathcal{F}_{\text{nearly-optimal}}\left(  \delta\right)  $)\textbf{. }

We recall that the goal in a quantum state discrimination problem is to
identify the actual state of a quantum system that is prepared, with a fixed
\emph{prior} probability, in a specific but unknown state that belongs to a
finite set of given possible states \cite{chefles00,croke09}. An essential
property of quantum mechanics is that two pure states cannot be distinguished
perfectly, if they are not orthogonal. Therefore, there are fundamental
limitations in quantum state discrimination protocols on the success with
which one can determine the actual state. Specifically, when the possible
states are not mutually orthogonal, it is not possible to develop a
state-distinguishing protocol that can discriminate between them perfectly.
Relevant state-distinguishing techniques are the unambiguous
\cite{ivanovic87,dieks88,peres88} and the ambiguous
\cite{holevo73,yuen75,helstrom76} state discrimination schemes. Unambiguous
state discrimination requires that, whenever a definite (that is, conclusive)
outcome is recorded after the measurement, the result must be error\ free
(that is, unambiguous). When the measurement fails to provide a definite
outcome, a nonzero probability of inconclusive outcomes has to be taken into
consideration. Then, the optimum unambiguous discrimination scheme is realized
when the failure probability (that is, the probability of inconclusive
outcomes) is minimum. On the contrary, ambiguous state discrimination requires
that a conclusive outcome has to be recorded in each single measurement. This
in turn implies that the discrimination is ambiguous since errors in the
conclusive result are unavoidable. Based on the measurement outcome, a guess
is made as to what the actual state of the quantum system was. Then, the
optimum ambiguous discrimination scheme is obtained when the error probability
(that is, the probability of making a wrong guess) is minimum. For a very
instructive discussion on how to interpret the original Farhi-Gutmann analog
quantum search algorithm as a method for improving the distinguishability of a
set of Hamiltonians by adding a controlled driving term, we refer to Ref.
\cite{childs00}.

Being in the framework of ambiguous quantum state discrimination, assume that
we are given one of two normalized quantum states $\left\vert \psi
_{1}\right\rangle =\left\vert w\right\rangle $ and $\left\vert \psi
_{2}\right\rangle =\left\vert \tilde{w}\right\rangle $, not necessarily
orthogonal, with \emph{prior} probabilities $p_{1}=p_{w}$ and $p_{2}%
\overset{\text{def}}{=}1-p_{1}=p_{\tilde{w}}$, respectively. Furthermore,
assume we have been asked to optimally determine which state we have actually
been given. It happens that the\textbf{ }minimum error probability
$p_{E}\left(  \delta\right)  $ is given by \cite{chefles00,croke09},%
\begin{equation}
p_{E}\left(  \delta\right)  =\frac{1}{2}\left(  1-\sqrt{1-4p_{w}p_{\tilde{w}%
}\cos^{2}\left(  \delta\right)  }\right)  \text{.} \label{me}%
\end{equation}
From Eq. (\ref{me}), we note that the minimum error probability $p_{E}$ is a
monotonic increasing function of the quantum mechanical overlap $\left\langle
w|\tilde{w}\right\rangle \overset{\text{def}}{=}\cos\left(  \delta\right)  $.
In particular, when $\delta$ approaches zero (that is, very large quantum
overlap) and $p_{w}=p_{\tilde{w}}$, it can become quite challenging to
distinguish the state $\left\vert \tilde{\omega}\right\rangle $ from the state
$\left\vert \omega\right\rangle $ since the probability of making an incorrect
guess approaches $50\%$.\textbf{ }Alternatively, we point out that in the
presence of large asymmetry between the prior probabilities $p_{w}$ and
$p_{\tilde{w}}$, the success of the discrimination protocol increases and, as
a consequence, it becomes less difficult to discriminate between $\left\vert
w\right\rangle $ and $\left\vert \tilde{w}\right\rangle $. For instance, when
$p_{\tilde{w}}=10^{-2}p_{w}$, the minimum error probability $p_{E}$ approaches
the liming value $10^{-2}$ as the angle $\delta$ approaches zero. We observe
that the quantum mechanical overlap $\left\langle w|\tilde{w}\right\rangle $
plays a key role in both quantum state discrimination and nearly optimal
analog quantum search. In the former case, it is the essential quantity that
sets a bound to the effectiveness of the discrimination scheme. In the latter
case, it quantifies the departure from the perfect unit fidelity.\textbf{ }In
view of the above mentioned considerations, it becomes useful to define an
approximate target state $\left\vert \tilde{w}\right\rangle $ as a quantum
state with a nearly optimal fidelity $\mathcal{F}_{\text{nearly-optimal}%
}\left(  \delta\right)  \overset{\text{def}}{=}\cos^{2}\left(  \delta\right)
$ that departs from the unit perfect fidelity $\mathcal{F}_{\text{optimal}%
}\overset{\text{def}}{=}1$ by an amount $\Delta\mathcal{F}\left(
\delta\right)  \overset{\text{def}}{=}\mathcal{F}_{\text{optimal}}%
-\mathcal{F}_{\text{nearly-optimal}}\left(  \delta\right)  $. We impose that
the quantity $\Delta\mathcal{F}\left(  \delta\right)  $ is smaller than the
minimum error probability $p_{E}\left(  \delta\right)  $ in Eq. (\ref{me}),%
\begin{equation}
0\leq\Delta\mathcal{F}\left(  \delta\right)  \overset{\text{def}}%
{=}\mathcal{F}_{\text{optimal}}-\mathcal{F}_{\text{nearly-optimal}}\left(
\delta\right)  \leq p_{E}\left(  \delta\right)  \text{.} \label{io1}%
\end{equation}
\begin{figure}[ptb]
\centering
\includegraphics[width=0.4\textwidth] {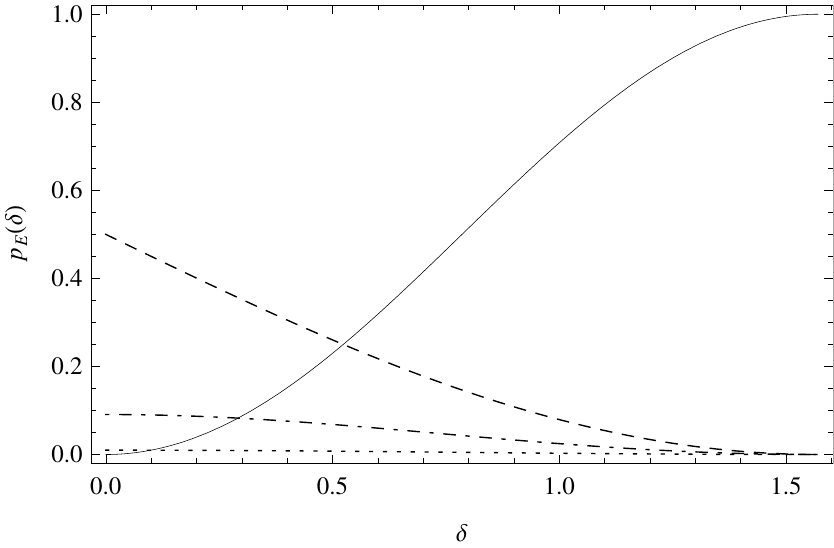}\caption{Plot of the minimum error
probability $p_{E}\left(  \delta\right)  $ v.s. the angle $\delta$ assuming
that $p_{w}=p_{\tilde{w}}$ (dashed line), $p_{w}=10p_{\tilde{w}}$ (dot-dashed
line), and $p_{w}=10^{2}p_{\tilde{w}}$ (dotted line). The thin line denotes
the deviation $\Delta\mathcal{F}$ from the optimal fidelity.}%
\label{fig3}%
\end{figure}Eq. (\ref{io1}) implies that we choose to select the lower bounds
of the nearly optimal fidelity values $\left\{  \mathcal{F}%
_{\text{nearly-optimal}}\left(  \delta\right)  \right\}  $ in such a manner
that their deviations $\left\{  \Delta\mathcal{F}\left(  \delta\right)
\right\}  $ from the unit fidelity $\mathcal{F}_{\text{optimal}}$ are less
than the minimum error probability $p_{E}\left(  \delta\right)  $
characterizing the optimum ambiguous discrimination scheme between the two
nonorthogonal quantum states $\left\{  \left\vert \tilde{\omega}\right\rangle
\text{, }\left\vert \omega\right\rangle \right\}  $ (that is, the approximate
and the exact target states, respectively) yielding the chosen nearly optimal
fidelity values. A plot of $p_{E}\left(  \delta\right)  $ as a function of the
angle $\delta$ for a variety of sets of \emph{a priori} probabilities appears
in Fig.\textbf{ }$3$. \ We point out that the proposed inequality in Eq.
(\ref{io1}) is reminiscent of the (sandwich) Fuchs-van de Graaf inequalities
relating the trace norm distance and the fidelity as suitable measures of
closeness for quantum states \cite{fuchs99}. Indeed, the trace norm distance
appears in the expression for the minimum error probability while the fidelity
appears in the equation for the transition probability in our discussion. For
the sake of completeness, we also remark that while Fuchs and van de Graaf
were only concerned with the problem of distinguishing between two quantum
states with equal\textbf{ }\emph{a priori}\textbf{ }probabilities, an
extension of their analysis to the case of arbitrary \emph{a priori}\textbf{
}probabilities can be found in Ref. \cite{mr14}. As a final consideration, we
emphasize that in the limiting case of equal \emph{a priori}\textbf{
}probabilities together with the conditions\textbf{ }$0\leq\delta\ll
1$\textbf{, }it is straightforward to verify that our inequality in Eq.
(\ref{io1}) can be obtained from the Fuchs-van de Graaf inequalities. In
particular, as explained below, the sharpness of the bounds that we propose in
Eq. (\ref{io1}) can be tuned by means of the degree of asymmetry that
specifies the \emph{a priori} probabilities of the two quantum states being distinguished.

Recalling Eq. (\ref{epsilon}), the problem was the determination of how small
the \emph{real} parameter $\varepsilon$ in Eq. (\ref{epsilon}) should be. We
are now in the condition to provide an answer to this issue. We note that,
after some straightforward algebra and using Eq. (\ref{40}), the inequality
constraint in\ Eq. (\ref{io1}) yields%
\begin{equation}
0\leq\delta\leq\delta_{\text{max}}\left(  \bar{p}_{w}\right)  \overset
{\text{def}}{=}\cos^{-1}\left[  \sqrt{1-\bar{p}_{w}\left(  1-\bar{p}%
_{w}\right)  }\right]  \text{.} \label{tetamax}%
\end{equation}
The quantity $\bar{p}_{w}$ in Eq. (\ref{tetamax}) denotes a fixed value of the
prior probability $p_{w}$. Knowing that $\bar{p}_{\tilde{w}}+$ $\bar{p}_{w}=1$
and setting $\bar{p}_{\tilde{w}}=\alpha\bar{p}_{w}$ with $\alpha$ being the
degree of asymmetry between the two \emph{a priori }probabilities $\bar
{p}_{\tilde{w}}$ and $\bar{p}_{w}$, Eq. (\ref{tetamax}) becomes%
\begin{equation}
0\leq\delta\leq\delta_{\text{max}}\left(  \alpha\right)  \overset{\text{def}%
}{=}\cos^{-1}\left[  \sqrt{1-\alpha\left(  1+\alpha\right)  ^{-2}}\right]
\text{.} \label{tetamax2}%
\end{equation}
We observe that in the maximally symmetric scenario, we have $\left(  \bar
{p}_{w}\text{, }\bar{p}_{\tilde{w}}\right)  =\left(  1/2\text{, }1/2\right)  $
(that is, $\alpha=1$). Instead, in the maximally asymmetric scenario, we have
$\left(  \bar{p}_{w}\text{, }\bar{p}_{\tilde{w}}\right)  =\left(  1\text{,
}0\right)  $ (that is, $\alpha\ll1$) or $\left(  \bar{p}_{w}\text{, }\bar
{p}_{\tilde{w}}\right)  =\left(  0\text{, }1\right)  $ (that is, $\alpha\gg
1$). Equating Eqs. (\ref{epsilon}) and (\ref{tetamax2}), we obtain a formal
expression for $\varepsilon$ in terms of the degree of asymmetry coefficient
$\alpha$,%
\begin{equation}
\varepsilon\left(  \alpha\right)  \overset{\text{def}}{=}\alpha\left(
1+\alpha\right)  ^{-2}\text{.} \label{epsilonal}%
\end{equation}
From Eq. (\ref{epsilonal}), we note that to get an $\varepsilon$ of the order
of $10^{-2}$ $\ll1$, one needs to have $\alpha$ of the order of $10^{2}%
$.\textbf{ }As previously mentioned, highly successful discrimination
protocols characterized by very small minimum error probability values can be
obtained when strong asymmetries between the prior probabilities $p_{w}$ and
$p_{\tilde{w}}$ occur. In such cases, the values of $\delta_{\text{max}%
}\left(  \bar{p}_{w}\right)  $ can be rather small. For instance, when
$p_{w}=10^{3}p_{\tilde{w}}$, $\delta_{\text{max}}\simeq3.16\times10^{-2}$ and
$\mathcal{F}_{\text{nearly-optimal}}\geq0.999$. We point out that this order
of magnitude of angles is quite small. Indeed, in real world settings one
necessarily deals with imperfect measurements where intrinsic uncertainties
yielding non-negligible systematic errors may be present. For instance, the
intrinsic uncertainty of a polarization rotor, typically of the order of
$2^{\mathrm{o}}-4^{\mathrm{o}}$ on the Bloch sphere (where $1^{\mathrm{o}%
}\simeq1.75\times10^{-2}$ rad.), limits the precision of the measurement on a
polarization qubit \cite{gisin12}.

For the sake of clarity, we point out that in our numerical calculations that
appear into the next section, the two \emph{a priori}\textbf{ }probabilities
$p_{w}$ and $p_{\tilde{w}}$ will not be arbitrarily chosen. They will be
chosen in such a manner to yield highly successful discrimination protocols in
which it is not difficult to discriminate between the exact and approximate
quantum states despite their high degree of overlap. Such protocols are
characterized by very low minimum error probability, which, in turn, is
achievable when there is a sufficient degree of asymmetry $\alpha$ between the
two\textbf{ }\emph{a priori}\textbf{ }probabilities corresponding to the two
states that we wish to distinguish (see Fig.\textbf{ }$3$). Then, to a larger
$\alpha$ there corresponds a smaller $\varepsilon$. Specifically, we shall be
considering asymmetries of the order of $10^{2}$ since they yield minimum
error probability values of the order of $10^{-2}$ which are low enough for
high degree of overlaps specified by angles of the order of $10^{-2}$ rad.
Then, being in the framework of highly successful quantum discrimination
protocols, we choose the accuracy the search has to achieve by imposing that
the departure\textbf{ }$\Delta\mathcal{F}$\textbf{ }of the unit
fidelity\textbf{ }$\mathcal{F}_{\text{optimal}}$\textbf{ }from the nearly
optimal desired fidelity value\textbf{ }$\mathcal{F}_{\text{nearly-optimal}}$
must be smaller than the already small minimum error probability $p_{E}$.
Imposing that $\delta_{\max}$ is of the order of the maximum achievable
angular resolution in a typical quantum mechanical experiment \cite{gisin12},
we find how asymmetric the two \emph{a priori} probabilities must be chosen in
order to guarantee the required accuracy the search should achieve.

In what follows, we compare the performances of the original and the modified
Farhi-Gutmann analog quantum search algorithms. Clearly, both algorithms are
stopped when they reach the same chosen lower bound.

\section{Comparison of the two search algorithms}

\begin{figure}[ptb]
\centering
\includegraphics[width=0.35\textwidth] {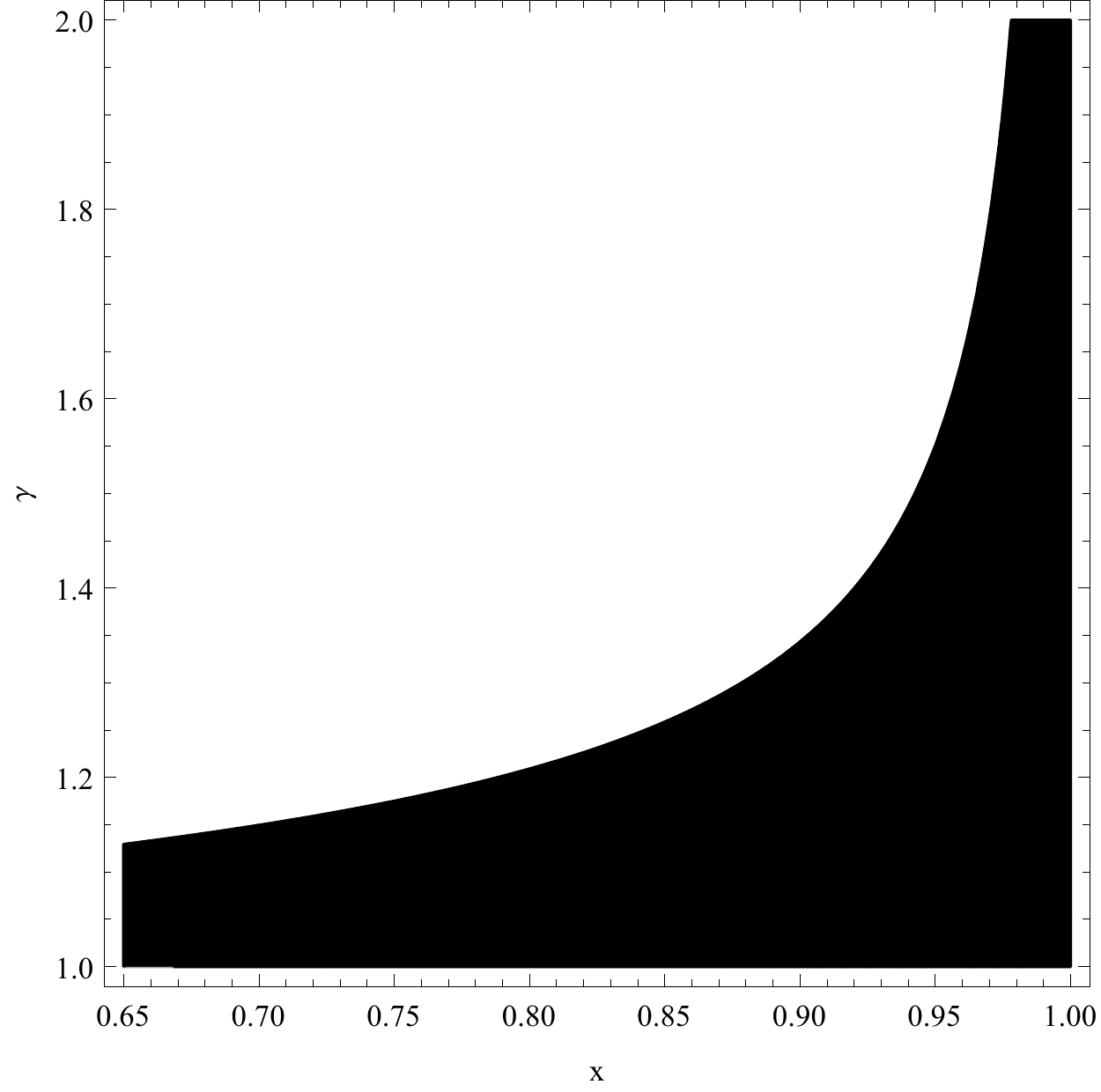}\caption{Plot of the
two-dimensional parametric region $r_{P}\left(  x\text{, }\gamma\right)  $
where the general algorithm outperforms the special algorithm in the working
assumption that $p_{w}=10^{2}p_{\tilde{w}}$ and $P_{\text{threshold}}%
=99.5\%$.}%
\label{fig4}%
\end{figure}In this section, we identify a two-dimensional parametric region
where the original algorithm is outperformed by the modified one provided one
focuses on nearly optimal fidelity values compatible with the maximum
resolution of the previously mentioned quantum discrimination protocol. For
the sake of convenience, we shall refer to the original Farhi-Gutmann
algorithm ($\gamma=1$) as the \emph{special} algorithm while the
\emph{general} algorithm denotes the modified version of the Farhi-Gutmann
algorithm where $E^{\prime}=\gamma E$ with $\gamma\geq1$. The transition
probability $\mathcal{P}\left(  t\text{; }x\text{, }E\text{, }E^{\prime
}\right)  $ with $E^{\prime}=\gamma E$ obtained for the general scenario
in\ Eq. (\ref{EF}) is\textbf{ }denoted as,%
\begin{equation}
\mathcal{P}_{\text{g}}\left(  t\right)  =\frac{x^{2}\left(  1+\gamma\right)
^{2}}{4x^{2}\gamma+\left(  1-\gamma\right)  ^{2}}\sin^{2}\left(  \frac
{E}{2\hslash}\sqrt{4x^{2}\gamma+\left(  1-\gamma\right)  ^{2}}t\right)
+x^{2}\cos^{2}\left(  \frac{E}{2\hslash}\sqrt{4x^{2}\gamma+\left(
1-\gamma\right)  ^{2}}t\right)  \text{.} \label{pg}%
\end{equation}
The maximum value of $\mathcal{P}_{\text{g}}\left(  t\right)  $ in Eq.
(\ref{pg}) is obtained at the time $\bar{t}_{\text{g}}$,%
\begin{equation}
\bar{t}_{\text{g}}\overset{\text{def}}{=}\frac{h}{4E}\frac{2}{\sqrt
{4x^{2}\gamma+\left(  1-\gamma\right)  ^{2}}}\text{,} \label{tg}%
\end{equation}
and, furthermore,%
\begin{equation}
\mathcal{P}_{\text{g}}^{\max}\overset{\text{def}}{=}\mathcal{P}_{\text{g}%
}\left(  \bar{t}_{\text{g}}\right)  =\frac{x^{2}\left(  1+\gamma\right)  ^{2}%
}{4x^{2}\gamma+\left(  1-\gamma\right)  ^{2}}\text{.} \label{gq}%
\end{equation}
Setting $\gamma=1$, we recover the main findings of Farhi and Gutmann.
Specifically, the transition probability $\mathcal{P}_{\text{g}}\left(
t\right)  $ in\ Eq. (\ref{pg}) reduces to $\mathcal{P}_{\text{s}}\left(
t\right)  \overset{\text{def}}{=}\mathcal{P}_{\text{Farhi-Gutmann}}\left(
t\text{; }x\text{, }E\right)  $ in Eq. (\ref{34}). The maximum value of
$\mathcal{P}_{\text{s}}\left(  t\right)  $ is obtained at the time $\bar
{t}_{\text{s}}$,%
\begin{equation}
\bar{t}_{\text{s}}\overset{\text{def}}{=}\frac{h}{4Ex}\text{,}%
\end{equation}
and, furthermore,%
\begin{equation}
\mathcal{P}_{\text{s}}^{\max}\overset{\text{def}}{=}\mathcal{P}_{\text{s}%
}\left(  \bar{t}_{\text{s}}\right)  =1\text{.}%
\end{equation}
We propose to rank the performance of the two algorithms in two different ways
given that Eq. (\ref{io1})\ is satisfied: i) rank the two algorithms by
comparing the two transitions probabilities evaluated at an identical
fixed-value of the travel time; ii) rank the two algorithms by comparing the
two minimum travel times needed to arrive at an identical fixed-value of the
transition probability.\begin{table}[t]
\centering
\par%
\begin{tabular}
[c]{c|c|c|c|c|c|c}\hline\hline
$x$ & $\delta$ $\left(  \times10^{-2}\right)  $ & $P_{\text{g}}^{\max}$ &
$\Delta\mathcal{F}$ $\left(  \times10^{-3}\right)  $ & $p_{E}$ $\left(
\times10^{-3}\right)  $ & $t_{\text{FG}}^{\left(  \text{special}\right)
}\left(  \times10^{-1}\right)  $ & $t_{\text{FG}}^{\left(  \text{general}%
\right)  }\left(  \times10^{-1}\right)  $\\\hline\hline
$0.65$ & $5.56$ & $0.9969$ & $3.100$ & $9.870$ & $3.67$ & $3.66$\\\hline
$0.70$ & $4.85$ & $0.9976$ & $2.400$ & $9.877$ & $3.42$ & $3.40$\\\hline
$0.75$ & $4.20$ & $0.9982$ & $1.800$ & $9.883$ & $3.20$ & $3.17$\\\hline
$0.80$ & $3.57$ & $0.9987$ & $1.300$ & $9.888$ & $3.01$ & $2.97$\\\hline
$0.85$ & $2.95$ & $0.9991$ & $0.900$ & $9.892$ & $2.84$ & $2.80$\\\hline
$0.90$ & $2.31$ & $0.9995$ & $0.500$ & $9.896$ & $2.68$ & $2.64$\\\hline
$0.95$ & $1.56$ & $0.9998$ & $0.200$ & $9.899$ & $2.55$ & $2.51$\\\hline
\end{tabular}
\caption{Illustrative numerical estimates of $t_{\text{FG}}^{\left(
\text{special}\right)  }$ and $t_{\text{FG}}^{\left(  \text{general}\right)
}$ for given values of $P_{\text{g}}^{\max}$ obtained by assuming $\gamma=1.1$
and $h=E=1$. For the selected values of the angle $\delta$, the numerical
estimates of $p_{E}$ are computed by considering the relation $p_{\omega
}=10^{2}p_{\tilde{\omega}}$ which yields $\delta_{\max}\simeq9.92\times
10^{-2}$.}%
\end{table}In the first scenario, since $\bar{t}_{\text{g}}\leq\bar
{t}_{\text{s}}$, we wish to determine whether or not there is, for a fixed
value of $\bar{t}_{\text{g}}$ less than $\bar{t}_{\text{s}}$, a
two-dimensional parametric region $\mathcal{R}_{t}\left(  x\text{, }%
\gamma\right)  $,%
\begin{equation}
\mathcal{R}_{t}\left(  x\text{, }\gamma\right)  \overset{\text{def}}%
{=}\left\{  \left(  x\text{, }\gamma\right)  \in\left(  0\text{, }1\right)
\times\left[  1\text{, }\infty\right)  :\mathcal{P}_{\text{g}}^{\max
}-\mathcal{P}_{\text{s}}\left(  \bar{t}_{\text{g}}\right)  >0\right\}
\text{,} \label{rpi1}%
\end{equation}
where the general algorithm outperforms the special algorithm in terms of
transition probability values in the working assumption of accepting
transition probability values less than one.

In the second scenario, since $\mathcal{P}_{\text{g}}^{\max}\leq
\mathcal{P}_{\text{s}}^{\max}$, we wish to determine whether or not there is,
for a fixed value of $\mathcal{P}_{\text{g}}^{\max}$ less than one, a
two-dimensional parametric region $\mathcal{R}_{\mathcal{P}}\left(  x\text{,
}\gamma\right)  $,%
\begin{equation}
\mathcal{R}_{\mathcal{P}}\left(  x\text{, }\gamma\right)  \overset{\text{def}%
}{=}\left\{  \left(  x\text{, }\gamma\right)  \in\left(  0\text{, }1\right)
\times\left[  1\text{, }\infty\right)  :\frac{\bar{t}_{\text{g}}}{\tilde
{t}_{\text{s}}}<1\right\}  \text{,} \label{rpi}%
\end{equation}
where the general algorithm outperforms the special algorithm in terms of
minimum travel times at fixed\textbf{ }nearly optimal transition probability values.

Note that $\tilde{t}_{\text{s}}$ in\ Eq. (\ref{rpi}) is such that,%
\begin{equation}
\mathcal{P}_{\text{s}}\left(  \tilde{t}_{\text{s}}\right)  =\mathcal{P}%
_{\text{g}}^{\max}\text{,}%
\end{equation}
where, after some algebraic manipulations, we find%
\begin{equation}
\tilde{t}_{\text{s}}\overset{\text{def}}{=}\frac{h}{2\pi Ex}\cos^{-1}\left[
\left(  \frac{1-\frac{x^{2}\left(  1+\gamma\right)  ^{2}}{4x^{2}\gamma+\left(
1-\gamma\right)  ^{2}}}{1-x^{2}}\right)  ^{\frac{1}{2}}\right]  \text{.}
\label{ts}%
\end{equation}
We emphasize that both two-dimensional parametric regions $\mathcal{R}%
_{t}\left(  x\text{, }\gamma\right)  $ and $\mathcal{R}_{\mathcal{P}}\left(
x\text{, }\gamma\right)  $ in Eqs. (\ref{rpi1}) and (\ref{rpi}), respectively,
are non-empty sets and in addition, it can be numerically verified that they
are identical. Moreover, for the sake of completeness we also point out that
one can investigate whether or not the general algorithm outperforms the
special one for a given transition probability threshold $\mathcal{P}%
_{\text{threshold}}$ given that Eq. (\ref{io1})\ is satisfied. For this
reason, we consider the following sub-region $r_{\mathcal{P}}\left(  x\text{,
}\gamma\right)  $ of $\mathcal{R}_{\mathcal{P}}\left(  x\text{, }%
\gamma\right)  $ in\ Eq. (\ref{rpi}),
\begin{equation}
r_{\mathcal{P}}\left(  x\text{, }\gamma\right)  \overset{\text{def}}%
{=}\left\{  \left(  x\text{, }\gamma\right)  \in\left(  0\text{, }1\right)
\times\left[  1\text{, }\infty\right)  :\frac{\bar{t}_{\text{g}}}{\tilde
{t}_{\text{s}}}<1\text{, and }\mathcal{P}_{\text{g}}^{\max}>\mathcal{P}%
_{\text{threshold}}\right\}  \text{.} \label{rpiccolo}%
\end{equation}
A plot of the two-dimensional parametric region $r_{P}\left(  x\text{, }%
\gamma\right)  $ in Eq. (\ref{rpiccolo}) where the general algorithm
outperforms the special algorithm when assuming $\mathcal{P}_{\text{threshold}%
}=99.5\%$ and $p_{w}=10^{2}p_{\tilde{w}}$ appears in Fig\textbf{.} $4$.

Finally, for the sake of clarity, we also present in Table I illustrative
numerical estimates of $\bar{t}_{\text{g}}\overset{\text{def}}{=}t_{\text{FG}%
}^{\left(  \text{general}\right)  }$ in Eq. (\ref{tg}) and $\tilde
{t}_{\text{s}}\overset{\text{def}}{=}t_{\text{FG}}^{\left(  \text{special}%
\right)  }$ in Eq. (\ref{ts}) for a number of selected values of the threshold
probability $\mathcal{P}_{\text{g}}^{\max}$ in Eq. (\ref{gq}) under the
working assumptions that $\gamma=1.1$ and $h=E=1$. Given the chosen values of
the angle $\delta$ in Table I, the numerical estimates of the minimum error
probability $p_{E}$ are calculated by assuming $p_{w}=10^{2}p_{\tilde{w}}$
which yields a maximum value of $\delta$ that equals $\delta_{\max}%
\simeq9.92\times10^{-2}$.

As a further clarifying remark, we point out that the numerical values
(originated from our proposed comparison of the two algorithms and reported in
Table I) were computed by considering the minimum time such that the desired
success probability values were obtained. We have simply chosen the desired
success probability values to be equal to the maximal success probability
values of the modified algorithm in our analysis. However, as evident from
Fig.\textbf{ }$5$, we could have chosen different threshold values (leading to
shorter minimum times) and still outperform the original algorithm (provided
that only a nearly optimal search is considered). As a side remark, we
emphasize that this line of reasoning extends naturally to any desired success
probability value taken as the maximal success probability\textbf{
}$\mathcal{P}_{\text{threshold}}=\mathcal{P}_{\text{g}}^{\max}$\textbf{
}with\textbf{ }$\mathcal{P}_{\text{g}}^{\max}=\mathcal{P}_{\text{g}}^{\max
}\left(  x\text{, }\gamma\right)  $ in Eq. (\ref{gq}) and where the
points\textbf{ }$\left\{  \left(  x\text{, }\gamma\right)  \right\}  $\textbf{
}can be chosen to belong to the black-colored region in Fig.\textbf{ }$4$.
Specifically, in Fig.\textbf{ }$5$ we plot $\mathcal{P}_{g}\left(  t\right)  $
(dashed line) and\textbf{ }$\mathcal{P}_{s}\left(  t\right)  $\textbf{ }(solid
line) as a function of time\textbf{ }$t$\textbf{. }For the sake of reasoning,
we set\textbf{ }$h=1$\textbf{, }$E=1$\textbf{, }$x=0.80$\textbf{, }and\textbf{
}$\gamma=1.1$\textbf{. }The two horizontal lines in Fig. $5$ denote two
selected success probability values\textbf{, }$\mathcal{P}_{\text{threshold}%
}^{\max}=0.9987$\textbf{ }(dotted line) and\textbf{ }$\mathcal{P}%
_{\text{threshold}}=0.9$\textbf{ }(dotted-dashed line). The threshold
value\textbf{ }$0.9987$\textbf{ }was taken from the fourth line in Table
I.\textbf{ }We note that the minimum time such that the two desired success
probability values are obtained is smaller in the case of the modified
algorithm for both selected thresholds. Specifically, in the first and second
scenarios, we have\textbf{ }$1.93\times10^{-1}$\textbf{ }$\leq2.02\times
10^{-1}$\textbf{ }and $2.97\times10^{-1}$\textbf{ }$\leq3.01\times10^{-1}%
$\textbf{,} respectively. Note that in the MKSA unit system,\textbf{ }$\left[
t\right]  _{\text{MKSA}}=\left[  h\right]  _{\text{MKSA}}\left[  E\right]
_{\text{MKSA}}^{-1}$\textbf{. }Since we have considered $E=1=h$\textbf{ }in
our numerical computations, time\textbf{ }$t$\textbf{ }is assumed to be dimensionless.

\begin{figure}[t]
\centering
\includegraphics[width=0.4\textwidth] {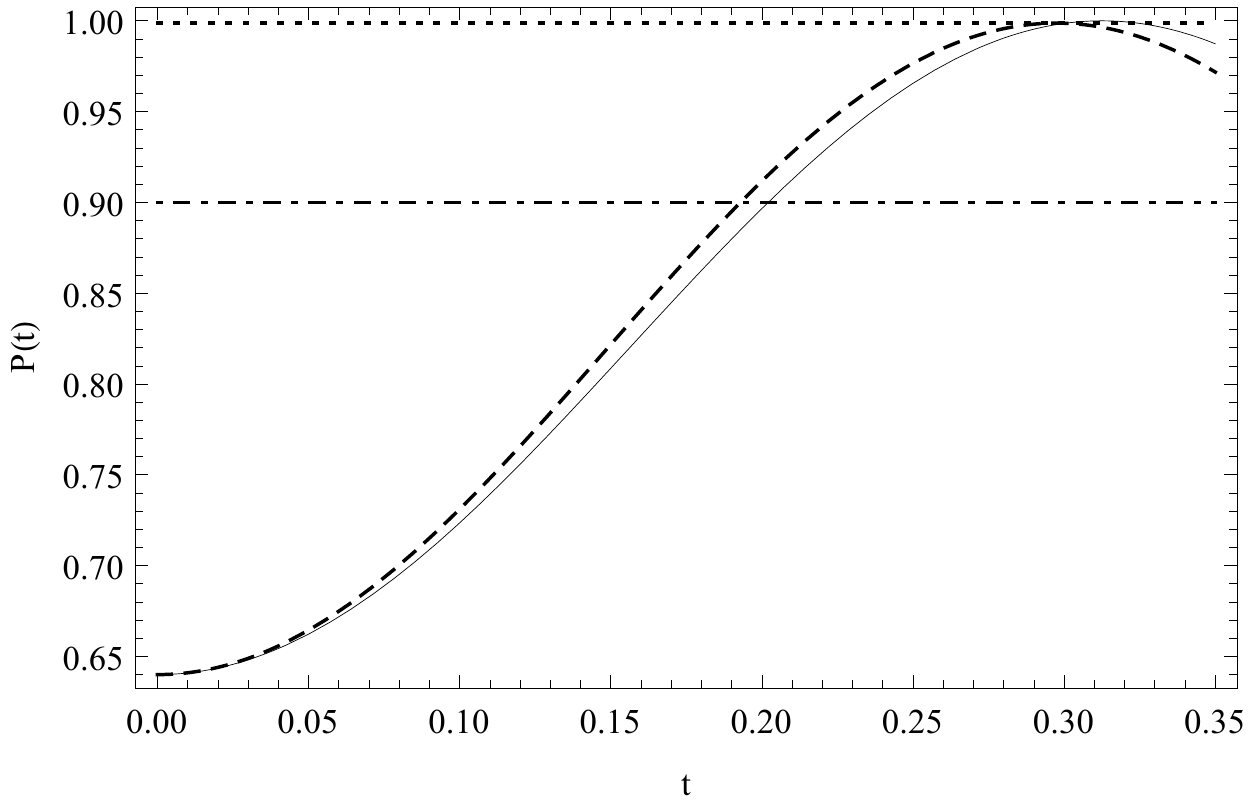}\caption{{Plot of $\mathcal{P}%
_{g}\left(  t\right)  $ (dashed line) and $\mathcal{P}_{s}\left(  t\right)  $
(solid line) versus $t$. We set $h=1$, $E=1$, $x=0.80$, and $\gamma=1.1$. The
two horizontal lines denote two selected success probability values,
$\mathcal{P}_{\text{threshold}}^{\max}=0.9987$ (dotted line) and
$\mathcal{P}_{\text{threshold}}=0.9$ (dot-dashed line). Observe that the
minimum time such that the two desired success probability values are obtained
is smaller in the case of the modified algorithm for both selected thresholds.
Specifically, in the first and second scenarios, we have $1.93\times10^{-1}$
$\leq2.02\times10^{-1}$ and $2.97\times10^{-1}$ $\leq3.01\times10^{-1}$,
respectively.}}%
\label{fig5}%
\end{figure}Finally, for a more detailed discussion on the probability of
small and large numerical values of the quantum overlap $x$, we refer to
Appendix A.

\section{Concluding Remarks}

In this paper, we focused on the problem of nearly optimal state searching and
we demonstrated that it is possible to modify the original Farhi-Gutmann
search Hamiltonian (see Eq. (\ref{hamilton2})) in order to speed up the
procedure for finding a suitably distributed unknown normalized quantum state
$\left\vert w\right\rangle $ under the working assumption that only a nearly
optimal fidelity is achieved,
\begin{equation}
\mathcal{P}_{\max}\left(  \delta\right)  \overset{\text{def}}{=}\cos
^{2}\left(  \delta\right)  \text{,}%
\end{equation}
with\textbf{ }$0\leq$\textbf{ }$\delta\overset{\text{def}}{=}\cos^{-1}\left[
\left\langle w|\tilde{w}\right\rangle \right]  \ll1$\textbf{ }where\textbf{
}$\left\vert w\right\rangle $\textbf{ }and\textbf{ }$\left\vert \tilde
{w}\right\rangle $\textbf{ }denote the exact and the approximate target
states, respectively. More specifically, upon\textbf{ }relaxing the working
assumptions of exact state overlap and uniform distribution of the target
state on the unit sphere in the $N$-dimensional \emph{complex} Hilbert space,
we showed that the proposed algorithm (see Eq. (\ref{hamilton2})) can indeed
outperform the original analog counterpart of a quantum search algorithm. This
enhanced\textbf{ }performance occurs for a convenient choice of both the ratio
$\gamma\overset{\text{def}}{=}E^{\prime}/E$ between the eigenvalues of the
modified search Hamiltonian and the quantum mechanical overlap $x\overset
{\text{def}}{=}\left\langle w|s\right\rangle $ between the initial $\left(
\left\vert s\right\rangle \right)  $ and the target $\left(  \left\vert
w\right\rangle \right)  $ states.

In summary, the main results of our work can be described as follows. First,
we extended the optimality proof presented in\ Ref. \cite{farhi98} to the case
of nearly optimal state searching and determined that the minimum time
interval to find the target state with probability $\mathcal{P}_{\max}\left(
\delta\right)  \overset{\text{def}}{=}\cos^{2}\left(  \delta\right)  $ is
given by (see Eq. (\ref{findingone})),
\begin{equation}
\tilde{t}_{\min}^{\left(  \text{approximate}\right)  }\overset{\text{def}}%
{=}\frac{\hslash}{2E}\left(  1-\delta\right)  \sqrt{N}\text{.}%
\end{equation}
Second, we computed the transition probability $\mathcal{P}\left(  t\text{;
}x\text{, }E\text{, }E^{\prime}\right)  $ (see Eq. (\ref{EF})) that arises
from our proposed search algorithm and determined its maximum value
$\mathcal{P}_{\max}\left(  x\text{, }\gamma\right)  $ (see Eq. (\ref{DDD}) and
Fig. $1$) together with the instant $t^{\ast}\left(  x\text{, }\gamma\right)
$ (see Eq. (\ref{start})) at\textbf{ }which this maximum is achieved. Third,
from the formal introduction of the quantity $\delta$ in Eq. (\ref{imperfect})
and the expression of $\mathcal{P}_{\max}\left(  x\text{, }\gamma\right)  $ in
Eq. (\ref{DDD}), we found an explicit expression for the quantity
$\delta=\delta\left(  x\text{, }\gamma\right)  $ (see Eq. (\ref{deltaequation}%
) and Fig. $2$) responsible for our imperfect state search in terms of the two
parameters $x$ and $\gamma$ that characterize our search Hamiltonian,
\begin{equation}
\delta\left(  x\text{, }\gamma\right)  \overset{\text{def}}{=}\cos^{-1}\left[
\frac{\left(  1+\gamma\right)  x}{\sqrt{\left(  1-\gamma\right)  ^{2}+4\gamma
x^{2}}}\right]  \text{.} \label{deltarel}%
\end{equation}
Fourth, we introduced the notion of approximate target state $\left\vert
\tilde{w}\right\rangle $. Such a quantum state $\left\vert \tilde
{w}\right\rangle $ is characterized by a sufficiently large overlap with the
exact target state $\left\vert w\right\rangle $ so that, as we proposed, the
deviation $\Delta\mathcal{F}\left(  \delta\right)  $ from the unit fidelity
$\mathcal{F}=1$ is smaller than the minimum error probability $p_{E}\left(
\delta\right)  $ achievable in an optimal quantum state discrimination
protocol (see Fig. $3$) for the set of quantum states $\left\{  \left\vert
w\right\rangle \text{, }\left\vert \tilde{w}\right\rangle \right\}  $,%
\begin{equation}
0\leq\Delta\mathcal{F}\left(  \delta\right)  =1-\cos^{2}\left(  \delta\right)
\leq p_{E}\left(  \delta\right)  \text{,}%
\end{equation}
with $p_{E}\left(  \delta\right)  $ defined in\ Eq. (\ref{me}). Finally, using
numerical methods, we showed that for a fixed transition probability threshold
$0\leq\mathcal{P}_{\text{threshold}}<1$ and suitably chosen non-uniform
probability distribution functions of the target state on the unit sphere (see
Eq. (\ref{ass})), there exist two-dimensional parametric regions $r_{P}\left(
x\text{, }\gamma\right)  $ (see Eq. (\ref{rpiccolo})) in which the proposed
algorithm outperforms the original one in terms of speed (see Fig. $4$, Table
I, and Fig. $5$) for the special task (namely, finding the approximate target
state) it sets out to solve.

We believe that our analysis presented in this paper can serve as a relevant
starting point for a more rigorous investigation that would include both
theoretical and experimental aspects of the tradeoff between fidelity and run
time of quantum search algorithms. For instance, it is known that to mitigate
the effect of decoherence originating from the interaction of a quantum system
with the environment, it is helpful to decrease the control time of the
control fields employed to generate a target quantum state or a target quantum
gate. Simultaneously, to enhance the fidelity of generating such targets and
reach values arbitrarily close to the maximum $\mathcal{F}=1$, it may be
convenient to increase the control time beyond a certain critical value. When
the control time reaches a certain value that may be close to the just
mentioned critical value however, decoherence can become a dominant effect.
Therefore, investigating the tradeoff between fidelity and time control can be
of great practical importance in quantum computing
\cite{rabitz12,rabitz15,cappellaro18}.\ For instance, one can design
algorithms seeking suboptimal control solutions for much reduced computational
effort, since it is very challenging to find a rigorous optimal time control
and in many cases the control is only required to be sufficiently precise and
short. For example, the fidelity of tomography experiments is rarely above
$99\%$ due to the limited control precision of the tomographic experimental
techniques as pointed out in Ref. \cite{rabitz15}. Under such conditions, it
is unnecessary to prolong the control time since the departure from the
optimal scenario is essentially negligible. Hence, it can certainly prove
worthwhile to design slightly suboptimal algorithms that can be much cheaper
computationally. Our analysis can be improved in a number of ways. First, our
work can be strengthened by explicitly considering the quantum measurement
process in our analysis. For instance, it is known that quantum measurement
can be regarded as a decoherence process that potentially enhance or reduce a
transition probability between two quantum states \cite{fritz10}. Second, as a
starting point of our investigation, we could consider a more general
time-independent search Hamiltonian where the probability of obtaining the
target state can be very large and nearly one \cite{kwon02}. Furthermore, in
order to work in a more realistic setting, nearly optimal time-dependent
search Hamiltonians specified in terms of the so-called schedule function of
the search algorithm could also be considered \cite{chuang17}. Finally, from a
more ambitious perspective, we could explore the manner in which recent
investigations on intelligent forms of quantum searching algorithms in the
presence of imperfections, with the help of techniques borrowed from quantum
machine learning \cite{lloyd17}, would connect to our nearly optimal quantum
search problem.

In conclusion,\textbf{ }we hope to pursue a more rigorous and realistic
analysis that includes these intriguing theoretical and experimental aspects
in forthcoming scientific efforts.

\begin{acknowledgments}
C. C. is grateful to the United States Air Force Research Laboratory (AFRL)
Summer Faculty Fellowship Program for providing support for this work. Any
opinions, findings and conclusions or recommendations expressed in this paper
are those of the authors and do not necessarily reflect the views of AFRL.
Finally, constructive criticism from two anonymous referees leading to an
improved version of this manuscript are sincerely acknowledged by the authors.
\end{acknowledgments}

\bigskip\pagebreak

\appendix

\section{Numerical values of the quantum overlap}

In this Appendix, we present a more in depth discussion on the probability of
small and large numerical values of the quantum overlap $x$.

In general, one may argue that large values of the overlap $x$ are not very
typical or particularly realistic since they occur when the source state is
essentially already the desired target state. This objection has its merit
when one assumes that the realistic scenario is the one in which no \emph{a
priori} information on the target state is available and, in particular, the
target state is assumed to be a normalized vector in a list of $N$ possible
orthonormal states. Indeed, within these working assumptions, $x=N^{-\frac
{1}{2}}$ and the largest value of $x$ that would be considered realistic is
obtained for $N=4$ and equals $x=1/2$. However, one can envision a
circumstance in which \emph{a priori }relevant\emph{\ }knowledge is available
so that the target state is a suitably chosen non-uniformly distributed
normalized vector in a $N$-dimensional vector space. In such a scenario, the
probability of occurrence of large values of $x$ can begin to be
non-negligible. Thus, large values of $x$ can become realistic in these new
working conditions. We point out that \emph{a priori }knowledge happens to be
useful in both quantum metrology and imperfect (optimal) quantum cloning. In
the former case, in order to design the best parameter estimation scheme, some
form of \emph{a priori} knowledge of such a parameter is required
\cite{paris18}. In the latter case, instead, knowledge of the distribution of
qubits on the Bloch sphere that encodes some \emph{a priori} information on a
given quantum state that one desires to clone is taken advantage of in order
to achieve the highest fidelity \cite{karol10,kang16}. For instance, in the
specific framework of phase-independent quantum cloning, one assumes to clone
qubits that are \emph{a priori} known to be symmetrically distributed around
the Bloch vector \cite{karol10}.

As pointed out in Section II, the optimality proof by Farhi and Gutmann occurs
under the special working assumption that the target state $\left\vert
w\right\rangle $ is an unknown element of a given orthonormal basis $\left\{
\left\vert a\right\rangle \right\}  $ with $1\leq a\leq N$ of an
$N$-dimensional complex Hilbert space $\mathcal{H}_{2}^{n}$ with
$N\overset{\text{def}}{=}2^{n}$. As pointed out earlier, in this scenario it
is rather unlikely that $x=N^{-\frac{1}{2}}$ assumes values close to one. In
their more general setting however, Farhi and Gutmann assume that the target
state $\left\vert w\right\rangle $ is an arbitrary normalized vector in a
$N$-dimensional complex Hilbert space which is \emph{uniformly} distributed on
the unit sphere \cite{farhi98}.

We briefly recall that the space enclosed by a $\left(  2N-1\right)
$-dimensional unit sphere $\mathcal{S}^{2N-1}$ is a $(2N)$-ball whose
infinitesimal volume element in spherical coordinates is given by,%
\begin{equation}
dV_{2N\text{-ball}}^{\left(  \text{spherical}\right)  }\overset{\text{def}}%
{=}r^{2N-1}\sin^{2N-2}\left(  \theta_{1}\right)  \sin^{2N-3}\left(  \theta
_{2}\right)  ...\sin\left(  \theta_{2N-2}\right)  drd\theta_{1}d\theta
_{2}...d\theta_{2N-2}d\theta_{2N-1}\text{,} \label{nball}%
\end{equation}
where $\theta_{i}\in\left[  0\text{, }\pi\right)  $ for any $1\leq i\leq2N-2$
and $\theta_{2N-1}\in\left[  0\text{, }2\pi\right)  $. Furthermore, the volume
element of the $\left(  2N-1\right)  $-dimensional unit sphere generalizes the
concept of the area element of a two-dimensional unit sphere and, from Eq.
(\ref{nball}), is given by%
\begin{equation}
dV_{\mathcal{S}^{2N-1}}^{\left(  \text{spherical}\right)  }\overset
{\text{def}}{=}\sin^{2N-2}\left(  \theta_{1}\right)  \sin^{2N-3}\left(
\theta_{2}\right)  ...\sin\left(  \theta_{2N-2}\right)  drd\theta_{1}%
d\theta_{2}...d\theta_{2N-2}d\theta_{2N-1}\text{.} \label{area}%
\end{equation}
Returning to our discussion, we assume that the probability density function
$\rho_{w}\left(  \theta\right)  $ that describes the distribution of the
states $\left\vert w\right\rangle $ on the $\left(  2N-1\right)  $-dimensional
unit sphere depends on the coordinate $\theta_{1}$, denoted as $\theta$, which
is uniform with respect to the remaining coordinates. Therefore, by\textbf{
}marginalizing over all the unimportant integration variables but $\theta$
with $x\overset{\text{def}}{=}\left\vert \left\langle s|w\right\rangle
\right\vert =\cos\left(  \theta\right)  $ where $0\leq\theta\leq\pi/2$, we
find that the probability that $x$ is greater than a given value $\bar{x}$ is
given by \cite{farhi98},
\begin{equation}
\text{\textrm{Prob}}\left(  x\geq\bar{x}\right)  \overset{\text{def}}{=}%
\frac{\int_{0}^{\cos^{-1}\left(  \bar{x}\right)  }\rho_{w}\left(
\theta\right)  \left[  \sin\left(  \theta\right)  \right]  ^{2N-2}d\theta
}{\int_{0}^{\frac{\pi}{2}}\rho_{w}\left(  \theta\right)  \left[  \sin\left(
\theta\right)  \right]  ^{2N-2}d\theta}\text{.} \label{prob}%
\end{equation}
The quantity $\rho_{w}\left(  \theta\right)  $ in Eq. (\ref{prob}) denotes a
well-defined probability density function (pdf), that is to say, a pdf that is
positive and normalized to one. For the sake of reasoning, we select
$\left\vert s\right\rangle $ to be at the north pole. Generalizations to less
peculiar scenarios are straightforward. In what follows, we provide some
rationale for our choice of $\rho_{w}\left(  \theta\right)  $. The functional
form of $\rho_{w}\left(  \theta\right)  $ is essentially that of a Gaussian
with mean $\mu_{\theta}$ and variance $\sigma_{\theta}^{2}$ multiplied by a
suitably chosen oscillatory function. The mean is set equal zero, while any
value of $\mu_{\theta}$ between $0$ and $\pi/2$ can be chosen provided that
the variance $\sigma_{\theta}^{2}$ is not too small. Furthermore, the
multiplying factor in the proposed expression of $\rho_{w}\left(
\theta\right)  $ is chosen in such a manner as\textbf{ }to substantially
mitigate the oscillatory behavior of $\left[  \sin\left(  \theta\right)
\right]  ^{2N-2}$ in Eq. (\ref{prob}) and leads when multiplied with it, to an
approximately constant function over the selected domain of integration.
Practically, one can consider a narrowly distributed Gaussian peaked nearby
the location of the initial state or, for a Gaussian peaked far away from such
a location, the width of the Gaussian has to be suitably larger. Under this
assumption, a convenient choice for our analysis is given by the following
pdf,%
\begin{equation}
\rho_{w}\left(  \theta\right)  \overset{\text{def}}{=}\mathcal{N}\frac
{\exp\left(  -\frac{\theta^{2}}{2\sigma_{\theta}^{2}}\right)  }{1+\left[
10\sin\left(  \theta\right)  \right]  ^{2N-2}}\text{.} \label{ass}%
\end{equation}
In Eq. (\ref{ass}), $\mathcal{N}=\mathcal{N}\left(  N\text{, }\sigma_{\theta
}^{2}\right)  $ is a normalization factor that depends on the choice of $N$
and $\sigma_{\theta}^{2}$. For instance, for $N=16$, $\sigma_{\theta}^{2}=1$,
and $\bar{x}=0.95$, by means of numerical integration, we find \textrm{Prob}%
$\left(  x\geq0.95\right)  \simeq21\%$. We note that with a smaller
$\sigma_{\theta}^{2}$, the probability of $x$ being greater than a selected
value $\bar{x}$ increases and asymptotically approaches unity. For instance,
for $\sigma_{\theta}^{2}=10^{-1}$ and $\sigma_{\theta}^{2}=10^{-2}$, we
observe that \textrm{Prob}$\left(  x\geq0.95\right)  \simeq58\%$ and
\textrm{Prob}$\left(  x\geq0.95\right)  \simeq99\%$, respectively. We point
out that if $N=16$ and $\rho_{w}\left(  \theta\right)  $ is uniform as
selected in Ref. \cite{farhi98}, \textrm{Prob}$\left(  x\geq0.95\right)
\simeq3.2\times10^{-17}$. Therefore, although the probability is not exactly
zero, it is very unlikely that $x$ is close to $1$ when assuming uniformity
for $\rho_{w}\left(  \theta\right)  $. For this reason, we considered here
nearly optimal (imperfect) state searches where the target state is not
uniformly distributed on the unit sphere. As a final remark, we point out that
the assumption that the target state $\left\vert w\right\rangle $ is selected
at random means that we assume absolute ignorance (that is, maximum entropy)
about the location of the target. If we somehow learn an important piece of
information about the location of the target however, one can think of
updating his/her state of knowledge (see Ref. \cite{cafaropre}, for instance)
about the target with a new probability density function of the state on the
$\left(  2N-1\right)  $-dimensional space. As a consequence, the search
algorithm can be adapted to the target in order to improve the efficiency of
the searching scheme. We emphasize that these considerations are reminiscent
of what happens in channel-adapted quantum error correction
\cite{cafaro14,fletcher07,cafaroosid} and adaptive quantum computing
\cite{briegel15} where classical learning techniques can be used to enhance
the performance of certain quantum tasks \cite{briegel16}. We leave the
exploration of these intriguing ideas to future investigations.
\end{document}